\documentclass[11pt,english]{article}
\usepackage{lmodern}

\let\origrmdefault\rmdefault
\usepackage[math]{iwona}
\renewcommand{\rmdefault}{\origrmdefault}
\usepackage[T1]{fontenc}
\usepackage[latin9]{inputenc}
\usepackage{geometry}
\usepackage{array}
\usepackage{multirow}
\usepackage{amsmath}
\usepackage{amssymb}
\usepackage{graphicx}
\usepackage{setspace}
\usepackage[authoryear]{natbib}
\makeatletter

\providecommand{\tabularnewline}{\\}

\newcommand{\lyxaddress}[1]{
	\par {\raggedright #1
	\vspace{1.4em}
	\noindent\par}
}

\@ifundefined{date}{}{\date{}}
\usepackage{pdflscape}

\makeatother

\usepackage{babel}
\begin{document}
\title{Minimal chaotic models from the Volterra gyrostat}
\author{Ashwin K Seshadri\textsuperscript{1} and S Lakshmivarahan\textsuperscript{2}}
\maketitle

\lyxaddress{\textsuperscript{1}Centre for Atmospheric and Oceanic Sciences and
Divecha Centre for Climate Change, Indian Institute of Science, Bangalore
560012, India. Email: ashwins@iisc.ac.in.}

\lyxaddress{\textsuperscript{2}Emeritus faculty at the School of Computer Science,
University of Oklahoma, Norman, OK 73012, USA. Email: varahan@ou.edu.}

\subsection*{Declarations of interest: none}

\pagebreak{}
\begin{abstract}
Low-order models obtained through Galerkin projection of several physically
important systems (e.g., Rayleigh-Bénard convection, mid-latitude
quasi-geostrophic dynamics, and vorticity dynamics) appear in the
form of coupled gyrostats. Forced dissipative chaos is an important
phenomenon in these models, and this paper introduces and identifies
``minimal chaotic models'' (MCMs), in the sense of having the fewest
external forcing and linear dissipation terms, for the class of models
arising from an underlying gyrostat core. The identification of MCMs
reveals common conditions for chaos across a wide variety of physical
systems. It is shown here that a critical distinction is whether the
gyrostat core (without forcing or dissipation) conserves energy, depending
on whether the sum of the quadratic coefficients is zero. The paper
demonstrates that, for the energy-conserving condition of the gyrostat
core, the requirement of a characteristic pair of fixed points that
repel the chaotic flow dictates placement of forcing and dissipation
in the minimal chaotic models. In contrast if the core does not conserve
energy, the forcing can be arranged in additional ways for chaos to
appear in the subclasses where linear feedbacks render fewer invariants
in the gyrostat core. In all cases, the linear mode must experience
dissipation for chaos to arise. The Volterra gyrostat presents a clear
example where the arrangement of fixed points circumscribes more complex
dynamics.
\end{abstract}

\section{Introduction}

\subsection{Background and literature}

Forced dissipative chaos appears in many climate and geophysical flows
(\citet{Howard1986,Swart1988,Tong2009}), with many dynamical systems
combining effects of forcing as well as dissipation (\citet{Lorenz1960,Hide1994}).
A well-known example involves the special projection of Rayleigh-Bénard
convection onto $3$ modes, with one momentum and two thermal components
(\citet{Lorenz1963}). One of the earliest examples of chaos in $3$-dimensional
flows, the forcing in this model comes from an external thermal gradient
driving dynamics away from equilibrium, while dissipation appears
in both momentum and temperature dynamics. This model simplified an
earlier low-order model derived by Saltzman (\citet{Saltzman1962}),
and is derived from governing equations of convection in a fluid of
uniform depth forced by an external thermal gradient. Since the model
approximates incompressible flow in two dimensions, the equations
describe streamfunction evolution (in a single mode), in addition
to two temperature modes evolving nonlinearly (\citet{Lorenz1963}).
The route to chaos in this forced-dissipative model has been studied
extensively (\citet{Sparrow1982}) and involves a sequence of bifurcations
that are initiated by destabilization of a pair of fixed points, eventually
giving rise to a strange attractor.

Many other examples of nonlinear flows in climate, meteorology, and
geophysics, such as wave-mean flow interactions in mid-latitudes in
the context of quasi-geostrophic dynamics (\citet{Swart1988}), vorticity
dynamics (\citet{Lorenz1960,Charney1979}), convection in shear flows
(\citet{Howard1986,Thiffeault1996,Gluhovsky1999}), as well as flows
in electrically conducting fluids (\citet{Kennett1976,Hide1994}),
have yielded low-order models admitting complex evolution. The governing
equations in such systems have generally been discretized using the
Galerkin projection method (\citet{Holmes2012}). A general difficulty
with such model reductions has been that the resulting equations do
not necessarily retain the invariants of the governing equations in
the limit without any external forcing or dissipation (\citet{Gluhovsky2002,Gluhovsky2006,Thiffeault1996}).
Several authors have considered the difficulties that can arise if
the invariants are not held by truncated equations, and maintaining
such invariants is important to avoid nonphysical numerical dissipation,
preserve analogous energy transfers in the truncated equations, and
avoid spurious divergent solutions (\citet{Thiffeault1996}).

It has been shown that the Volterra gyrostat can naturally form the
building block of Galerkin projections of the governing partial-differential
equations, while maintaining energy conservation in the unforced and
dissipationless limit (\citet{Oboukhov1975,Thiffeault1996,Gluhovsky2002}).
Motivated by early studies pointing to the importance of modular approaches
to constructing low order models (LOMs) from the governing equations
(\citet{Oboukhov1975}), as well as the role of systematic approaches
for ensuring the maintenance of invariants within the conservative
core (\citet{Thiffeault1996,Gluhovsky1997,Gluhovsky1999}), more recent
studies have not only expanded on the earlier approaches but also
exemplified the ideas (\citet{Gluhovsky2006,Lakshmivarahan2006,Lakshmivarahan2008,Tong2008}).
Many examples from these domains have been identified that can be
described in terms of systems of coupled gyrostats (\citet{Gluhovsky1997,Gluhovsky1999,Gluhovsky2002,Gluhovsky2006,Tong2008}).
Furthermore, where there exist quadratic invariants such as kinetic
energy or the squared angular momentum, these are maintained in the
resulting truncated equations as well. Owing to such properties, it
is of widespread importance to study dynamics of models arising from
the Volterra gyrostat. An important generalization of the Volterra
gyrostat involves the inclusion of nonlinear feedback between modes
(\citet{Lakshmivarahan2008a}).The nature of dynamics in the presence
of gyrostatic forces is an active area of research (\citet{Amer2008,Amer2017}).

The Volterra gyrostat presents a three-dimensional volume-conserving
flow with a skew-symmetric structure of linear feedbacks and nonlinear
interactions between modes. To form the building blocks of LOMs, it
has been convenient to transform the original equations written by
Volterra through a smooth change of variables (\citet{Gluhovsky1999}).
As a result, in general, the building blocks of these LOMs have two
invariants, analogous to the conservation of kinetic energy and angular
momentum in the physical gyrostat. Each invariant confines the dynamics
to a two-dimensional surface, and their intersection gives rise to
oscillatory dynamics for the gyrostat core of these models. With inclusion
of forcing and dissipation (F\&D), there are no longer any quadratic
invariants, and thus F\&D can generate higher dimensional dynamics,
including chaos.

In the original gyrostat equations, the sum of the quadratic coefficients
is zero, a property rooted in the kinetic energy conservation of the
physical system (\citet{Gluhovsky1999}). It is remarkable that this
constraint leads to not one, but two, quadratic invariants in the
gyrostat core (\citet{Seshadri2023}). As a result, three-dimensional
systems where this constraint holds require forcing and dissipation
to be present for chaos to appear. More generally, a simple modification
to the gyrostat\textquoteright s quadratic coefficients (even without
F\&D being present) can reduce the number of quadratic invariants
(\citet{Seshadri2023}). In particular, it was previously shown that
if the sum of quadratic coefficients is nonzero, the gyrostat does
not conserve energy and then the number of invariants depends on the
number of linear feedbacks (\citet{Seshadri2023}). For example, if
there are three distinct linear feedbacks, then there are no quadratic
invariants in the gyrostat with a nonzero sum of quadratic coefficients,
and such models can admit chaotic dynamics even without F\&D (\citet{Seshadri2023}).

While maintaining other features of the gyrostatic models such as
conservation of volumes in phase space, such cores can also naturally
appear in Galerkin projections. Depending on the number of quadratic
and linear terms present, \citet{Gluhovsky1999} have shown that the
Volterra gyrostat can be specialized into nine different subclasses
by setting various combinations of parameters to zero. Different subclasses
constitute different linear and nonlinear interactions between modes
and together describe the different types of gyrostat cores. The number
of quadratic invariants in the gyrostat core can be expected to influence
the ways in which chaos can be produced due to F\&D effects.

\subsection{Formulation of the problem and Minimal Chaotic Models (MCMs)}

This paper considers chaos in models with forcing and dissipation
added to the equations of a single Volterra gyrostat having linear
feedback terms.

Especially, considering the combined effects of forcing and dissipation,
we identify minimal chaotic models derived from the Volterra gyrostat.
``Minimal chaotic models (MCMs)'' as defined in this paper are those
cases that are proper subsets, containing the necessary (but not all)
the forcing and dissipation terms among the chaotic cases. If a chaotic
case can admit chaos upon omitting any forcing or dissipation term,
it is not an MCM. MCMs are important because they point to the necessary
conditions for chaos to appear in these models. In this work, MCMs
are characterized by the equations in which nonzero forcing and dissipation
must appear. Since the individual equations describe the evolution
of different modes in a Galerkin projection, identifying MCMs localizes
where external forcing and internal dissipation must be included in
the corresponding physical systems that are being discretized.

By relating the physics of the underlying system to the structure
of the LOMs that can yield chaotic dynamics, MCMs can help understand
and interpret the physical conditions for chaos. Moreover, when the
MCMs arise from the same type of underlying model, such as the Volterra
gyrostat, we show that a common account of conditions for chaos in
such models appears. For example, it is shown here that MCMs from
the Volterra gyrostat must necessarily involve dissipation in linearly
evolving variables of the physical system. Thus, while identifying
MCMs does not diminish the importance of numerical simulation for
investigating chaos, searching for MCMs can anticipate common conditions
and advance understanding of physical possibilities, especially where
the appearance of chaos is limited to small regions of parameter space.

The search for MCMs is not only a useful mathematical artifice but
can also have direct physical application. Gyrostats play important
roles in mechanical control, such as satellite stabilization (\citet{Wittenburg1977,Hughes1986}).
If the controller becomes chaotic, the controlled system will not
behave well. Avoiding chaotic dynamics of the gyrostat becomes critical
to its successful operation (\citet{Leipnik1981,Amer2017,ElSabaa2022}).
As illustrated by this application of the gyrostat, identifying MCMs
can be important in classes of models for which boundaries of chaotic
and non-chaotic dynamics are critical. The fact that MCMs embody necessary
conditions for chaos can also give rise to design problems in a variety
of domains.

\subsection{Research contributions in context of previous literature}

Many studies have examined chaos in forced-dissipative models that
can take the form of systems of coupled gyrostats (\citet{Lorenz1963,Brindley1980,Gibbon1982,Howard1986,Swart1988,Gluhovsky1999,Hide1994,Reiterer1998,Huang2023,Thiffeault1996}).
Given the focus on individual systems, such studies have usually not
sought to omit individual forcing and dissipation terms to examine
whether chaos persists in ever more reduced models. A different body
of work has systematically searched for simple chaotic flows through
numerical simulation, without regard to their application, with simplicity
being defined by the number of distinct terms in the vector field
(\citet{Sprott1994,Sprott2010,Sprott2014}). Since such simple chaotic
flows can take any form, they often involve fewer terms (\citet{Sprott1994})
than the MCMs based on the gyrostat in the present work. Since this
literature on simple chaotic flows has not been confined to any single
class of models, there is no common structure underlying these models.
Extensive specializations of this literature have also been made,
for example to chaotic flows with specified symmetries (\citet{Sprott2014}),
possessing given geometries of fixed points (\citet{Jafari2013,Barati2016,Wang2020}),
and no fixed points at all (\citet{Jafari2013a}). However, none of
these studies have established MCMs within a given class, or otherwise
characterized those chaotic models containing proper subsets of all
the terms. There has been no systematic investigation of MCMs for
a single class of model before, involving gyrostats or any other classes
of model.

Given their ubiquity in Galerkin projections (\citet{Gluhovsky1997,Gluhovsky1999,Gluhovsky2006}),
MCMs based on the gyrostat core present an important class of chaos.
For such models based on the Volterra gyrostat, prior studies have
examined chaos with F\&D (\citet{Lorenz1963,Sparrow1982,Gibbon1982,Swart1988,Hide1994}),
as well as chaos without F\&D arising from absence of any invariants
due to lack of energy conservation (\citet{Seshadri2023}). For chaos
with F\&D, there has been no systematic examination of how the conditions
for placing F\&D for chaos to be admitted depend on whether the gyrostat
core conserves energy. This paper bridges all of these gaps, by closely
considering MCMs with F\&D, with and without energy conservation in
the gyrostat core, showing that this condition makes an important
difference and is critical to our account of MCMs from the Volterra
gyrostat.

\subsection{Approach and organization of the paper}

Section 2 outlines the models and methods used in the paper. Section
3 identifies chaotic cases and the MCMs by simulating large ensembles.
We first consider the possibility of chaos due to F\&D without the
energy conservation constraint, where we must distinguish different
cases by merely identifying those equations where nonzero F\&D must
arise. Given the three components of the vector field, we obtain a
possible $2^{6}=64$ cases for placement of F\&D, out of which $2\times2^{3}=16$
cases have either no forcing or no dissipation (or neither) in any
of the equations. We need only consider the remaining $48$ cases
for the presence of chaos, and simulate ensembles to sample the parameter
space for each of these cases. Upon listing all the chaotic cases,
we note that there exist cases that are proper subsets, containing
some (but not all) of the forcing and dissipation terms. These proper
subsets are defined as \textquotedblleft minimal chaotic models (MCMs)\textquotedblright .
We identify all the MCMs for the gyrostat having two nonlinear terms.
There could be more than one MCM, corresponding to distinct proper
subsets. The significance of these MCMs lies not only in their specific
arrangements of forcing and dissipation, but also in common features
across different subclasses of the gyrostat. Following this, we consider
the effects of whether energy is conserved in the gyrostat's core,
which influences where forcing must be placed for chaos to appear
in the equations. The MCMs are summarized in Section 3. Section 4
accounts for the MCMs, with and without energy conservation in the
gyrostat core, and examines common features that are observed across
subclasses. This is followed by a discussion of implications and open
questions in Section 5, and lastly a summary of main results in Section
6.

\section{Models and methods}

Volterra's equations for the gyrostat are (\citet{Gluhovsky1999})
\begin{align}
K_{1}^{2}\dot{y}_{1} & =\left(K_{2}^{2}-K_{3}^{2}\right)y_{2}y_{3}+h_{2}y_{3}-h_{3}y_{2}\nonumber \\
K_{2}^{2}\dot{y}_{2} & =\left(K_{3}^{2}-K_{1}^{2}\right)y_{3}y_{1}+h_{3}y_{1}-h_{1}y_{3}\nonumber \\
K_{3}^{2}\dot{y}_{3} & =\left(K_{1}^{2}-K_{2}^{2}\right)y_{1}y_{2}+h_{1}y_{2}-h_{2}y_{1}\label{eq:p1}
\end{align}
with $y_{i}$, $i=1,2,3$, being the angular velocity of the carrier
body, $K_{i}^{2}=I_{i}$ the principal moments of inertia of the gyrostat,
and $h_{i}$ the fixed angular momenta of the rotor relative to the
carrier. Defining a new set of variables $x_{i}$ through $K_{i}y_{i}=x_{i}$
and introducing a new set of parameters related to those appearing
above as
\begin{align}
p & =K_{2}^{2}-K_{3}^{2},q=K_{3}^{2}-K_{1}^{2},\textrm{and }r=K_{1}^{2}-K_{2}^{2}\nonumber \\
a & =K_{1}h_{1},b=K_{2}h_{2},\textrm{and }c=K_{3}h_{3}\label{eq:p2}
\end{align}
the above system is rendered as the transformed gyrostat equations
(\citet{Gluhovsky1999,Seshadri2023}) 
\begin{align}
x_{1}' & =px_{2}x_{3}+bx_{3}-cx_{2}\nonumber \\
x_{2}' & =qx_{3}x_{1}+cx_{1}-ax_{3}\nonumber \\
x_{3}' & =rx_{1}x_{2}+ax_{2}-bx_{1},\label{eq:p3}
\end{align}
which naturally appears in modular form in many LOMs. Here $'$ denotes
$d/ds$, while $\dot{}$ is $d/dt$, where $t=K_{1}K_{2}K_{3}s$.
Henceforth we shall work exclusively with the model in Eq. (\ref{eq:p3})
and denote time appearing there as $t$. We shall refer to Eq. (\ref{eq:p3})
as the gyrostat core. The resulting flow preserves volumes in phase
space, as the trace of its Jacobian is zero. Broadly we must distinguish
two types of conditions for the gyrostat core:
\begin{itemize}
\item With $p+q+r=0$ following Eq. (\ref{eq:p2}), the gyrostat core conserves
kinetic energy $E=\frac{1}{2}\sum_{i=1}^{3}K_{i}^{2}y_{i}^{2}$, as
well as squared angular momentum $M=\frac{1}{2}\sum_{i=1}^{3}\left(K_{i}^{2}y_{i}+h_{i}\right)^{2}$,
and the trajectories are oscillatory for all initial conditions (\citet{Seshadri2023}).
Skew-symmetry in the linear feedbacks plays an important role in energy
conservation (Appendix 1).
\item In contrast for $p+q+r\neq0$, where the model in Eq. (\ref{eq:p3})
does not have a direct analogue to the physical gyrostat, the number
of invariants depends on the number of nonzero linear coefficients
$\left(a,b,c\right)$, with zero, one, and two invariants for three,
two, or fewer nonzero coefficients respectively. This situation appears
in LOMs, such as the maximum simplification equations (\citet{Lorenz1960,Lakshmivarahan2006}).
Only with all of $a,b,c\neq0$ in the absence of energy conservation
does the gyrostat core admit chaos as a result of no invariants being
present (\citet{Seshadri2023}).
\end{itemize}
Despite the necessity of non-conservation of energy from $p+q+r\neq0$
for chaos in the gyrostat core without F\&D being present, such a
distinction is not usually made for chaos in the presence of F\&D.
This is probably because counterparts of Eq. (\ref{eq:p3}) with F\&D
are known to present chaos even when energy conservation is present
and the underlying gyrostat core exhibits periodic dynamics. In this
paper we will distinguish whether or not energy is conserved in the
gyrostat core, and show that it affects the MCMs that are obtained.

In this paper we integrate models with F\&D
\begin{align}
x_{1}' & =px_{2}x_{3}+bx_{3}-cx_{2}-\epsilon_{1}x_{1}+F_{1}\nonumber \\
x_{2}' & =qx_{3}x_{1}+cx_{1}-ax_{3}-\epsilon_{2}x_{2}+F_{2}\nonumber \\
x_{3}' & =rx_{1}x_{2}+ax_{2}-bx_{1}-\epsilon_{3}x_{3}+F_{3}\label{eq:p4}
\end{align}
with $\epsilon_{i}\geq0$, $i=1,2,3$ and $F_{j}\in\mathbb{R}$, $j=1,2,3$
for varying model parameters $\left(p,q,r,a,b,c\right)$. First, we
shall sample from the parameter space without imposing the constraint
$p+q+r=0$, to identify chaotic dynamics with sparse inclusion of
forcing and dissipation, with as few of $\left(\epsilon_{1},\epsilon_{2},\epsilon_{3},F_{1},F_{2},F_{3}\right)$
being nonzero as possible.

\citet{Gluhovsky1999} have identified special cases (``subclasses'')
of Eq. (\ref{eq:p3}), by specializing the quadratic and linear coefficients.
With the energy conservation constraint that is present throughout
their analysis, there must be at least two nonzero $p,q,r$ otherwise
the model is linear, so without loss of generality they assume that
$p,q\neq0$. Further restricting various combinations of linear coefficients
to be zero gives nine subclasses in addition to the general case with
nonzero parameters. We organize our analysis by subclasses for investigating
the role of F\&D in producing chaos, considering only those subclasses
with two (but not three) quadratic terms (subclasses $1-4$ of \citet{Gluhovsky1999},
with $r=0$ in Eq. (\ref{eq:p4})). These subclasses restrict individual
parameters as follows: subclass $1$ ($r=0,b=c=0$), subclass $2$
($r=0,c=0$), subclass $3$ ($r=0,b=0$) , and subclass $4$ ($r=0$).
These all have two quadratic terms, while subclasses $5-7$ have three
quadratic terms, giving rise to higher-degree equations for the steady
states.\footnote{We have also omitted the degenerate cases in subclasses $8-9$, as
defined by \citet{Gluhovsky1999}, for which the dynamics are two-dimensional.} Our focus on subclasses $1-4$ gives a cubic equation for the steady
states, as shown below, making them amenable to a common framework.
Yet, there are differences between them, owing to the different number
and arrangement of linear feedbacks, giving rise to different MCMs
that are explicated through this common framework.

For each of the subclasses $1-4$ in \citet{Gluhovsky1999}, we distinguish
different cases according to where forcing and dissipation arise.
Thus for each subclass, we can have either $\epsilon_{i}=0$ or $\epsilon_{i}>0$
and likewise $F_{j}=0$ or $F_{j}\neq0$, for $i=1,2,3$ and $j=1,2,3$,
for a total of $4^{3}=64$ cases that differ in whether each equation
has dissipation, forcing, both or neither. Eight of these cases have
no dissipation in the model and eight have no forcing, leaving us
with $48$ cases having nonzero forcing as well as dissipation, which
are evaluated for the possibility of chaos. Additionally, each case
has parameters $\left(p,q,a,b,c\right)$ as well as values of $\left(\epsilon_{i},F_{j}\right)$
that must be varied $\left(r=0\right)$. For each of these $48$ cases
of each of the $4$ subclasses with two nonlinear terms, we generate
a large (as much as $10$-dimensional) Latin hypercube sample to vary
model parameters $\left(p,q,a,b,c\right)$, forcing and dissipation
$\left(\epsilon_{i},F_{j}\right)$, and initial conditions $\left(x_{10},x_{20},x_{30}\right)$.
Where dissipation appears in more than one mode, i.e. $\epsilon_{i},\epsilon_{j}>0$
for $i\neq j$ it is assigned the same value $\epsilon_{i}=\epsilon_{j}=\epsilon$
in the ensemble, and likewise with the value of the forcing terms,
since our goal is to identify MCMs.

For each member of the sample, we integrate from the corresponding
initial condition $\left(x_{10},x_{20},x_{30}\right)$ a $12$-dimensional
system describing the state-variables in Eq. (\ref{eq:p3}) as well
as the evolution of the $3\times3$ initial condition sensitivity
matrix $\mathrm{D_{x_{0}}}\left(t\right)$ that allows us to evaluate
the $3$ Lyapunov exponents of the model. Lyapunov exponents are computed
using a standard approach, based on singular value decomposition (SVD)
of the matrix $\mathrm{M}\left(t\right)$ defined as $\mathrm{M}\left(t\right)=\mathrm{D_{x_{0}}}\left(t\right)^{T}\mathrm{D_{x_{0}}}\left(t\right)$
(\citet{Pikovsky2016}). The matrix $\mathrm{D_{x_{0}}\left(\mathit{t}\right)\in\mathbb{R}^{\mathit{3\times3}}}$
consists of elements $\mathrm{D_{x_{0}}^{\mathit{i,j}}\left(\mathit{t}\right)=}\partial x_{i}\left(t\right)/\partial x_{j0}$
describing forward sensitivities to perturbation in the initial condition.
This matrix is initialized to the $3\times3$ identity matrix at $t=0$.
We integrate in time for the state $\left(x_{1}\left(t\right),x_{2}\left(t\right),x_{3}\left(t\right)\right)$
as well as the nine elements of $\mathrm{D_{x_{0}}}\left(t\right)$,
from $t=0$ to $t=5000$.

For a given case, defined by placement of forcing and dissipation,
once the $3$ Lyapunov exponents are estimated from the SVD of matrix
$\mathrm{M}\left(t=5000\right)$, and the largest Lyapunov exponent
(LLE) is found for each of the samples, we identify the most unstable
sample among those whose LLE is positive. In doing so, we consider
only those samples whose evolution is bounded and for which the LLE
calculation does not diverge. Many combinations of forcing and dissipation
can give rise to unbounded dynamics even for these volume-contracting
flows with dissipation. In case there exists a most unstable sample
with bounded evolution and positive but finite LLE, that case of the
subclass is examined further for the presence of chaos. Specifically,
we plot the $3$-dimensional orbits, time-series of $x_{1}$, Poincar$\acute{\textrm{e}}$
sections, power spectra of $x_{1}$, and the evolution of LLE with
time $\lambda\left(t\right)$. Together these multiple lines of evidence
allow us to distinguish chaotic orbits from non-chaotic ones among
those where LLE is estimated to be positive. The detailed results
are plotted in the Supplementary Information (SI).

This procedure is repeated for each of the $48$ cases for all $4$
subclasses. A case is chaotic if it presents a sample (i.e., choice
of parameters and initial conditions) with bounded evolution and concurrent
lines of evidence for chaos: broadband power spectrum, Poincar$\acute{\textrm{e}}$
sections with fractal structure, and evolution of LLE to a positive
value with time, in addition to highly irregular time-series of $x_{1}$.
For a given subclass, after chaotic cases (among the possible $48$)
have been identified, minimal chaotic cases are found by inspection.
A minimal chaotic case has a proper subset of all the forcing and
dissipation terms. Only those chaotic cases that require all the forcing
or dissipation terms therein to be present in order for chaos to occur
are termed as MCMs.

The above calculations are repeated for each of the $4$ subclasses
with the energy conservation constraint in the gyrostat core being
introduced, i.e. $p+q+r=0$, by setting $q=-p$, since $r=0$ in each
of these subclasses.

\section{Identification of minimal chaotic models (MCMs)}

Chaotic cases are identified from among each of the $48$ possible
arrangements of nonzero forcing and dissipation (``cases''), with
each case consisting of $2000$ samples across model parameters, initial
conditions, and the values of forcing and dissipation coefficients,
arranged in a Latin hypercube. While such an empirical approach cannot
ensure that all cases admitting chaos are identified, each minimal
chaotic case appears to have been identified explicitly, as shown
below. Multiple lines of evidence have been used to identify chaotic
cases. As described in Section 2, we shortlist simulations having
a positive value of the largest Lyapunov exponent (LLE) to probe potential
cases admitting chaos, and deploy further lines of evidence to confirm
the appearance of chaos: inspection of the orbits and time-series
of $x_{1}$, Poincar$\acute{\textrm{e}}$ sections, whether time-series
(once transients have diminished) have a broadband power spectrum,
and evolution of the LLE towards positive values.

\subsection*{No energy conservation constraint, i.e., $p+q+r\protect\neq0$:}

For subclass 1 there are many cases with positive LLE ($17$ have
been found explicitly; see Supplementary Information (SI) Figures
1-5), but only some of these are chaotic by the various measures examined.
The chaotic cases have nonzero $\left(F_{1},\epsilon_{3}\right)$,
$\left(F_{1},\epsilon_{1},\epsilon_{3}\right)$, $\left(F_{1},\epsilon_{2},\epsilon_{3}\right)$,
$\left(F_{1},F_{2},\epsilon_{3}\right)$, $\left(F_{1},F_{3},\epsilon_{3}\right)$,
$\left(F_{1},F_{2},F_{3},\epsilon_{3}\right)$ and $\left(F_{1},F_{2},F_{3},\epsilon_{2},\epsilon_{3}\right)$.
Despite the diversity of chaotic orbits (Figure 1), inspection identifies
a unique MCM with nonzero $\left(F_{1},\epsilon_{3}\right)$, since
each of these cases has nonzero $F_{1}$ and $\epsilon_{3}$. For
each case the most chaotic version of the $2000$-member ensemble,
having the largest positive value of LLE, has been plotted to illustrate
the appearance of chaos (Figure 1).

Similarly, for subclass $2$, of the $23$ cases with positive LLE
(SI Figures 6-10), there are $12$ cases that are chaotic by all the
measures in Section 2. These cases (orbits in Figure 2), have nonzero
$\left(F_{1},\epsilon_{3}\right)$,$\left(F_{1},\epsilon_{1},\epsilon_{3}\right)$,$\left(F_{1},\epsilon_{2},\epsilon_{3}\right)$,
$\left(F_{2},\epsilon_{3}\right)$,$\left(F_{2},\epsilon_{1},\epsilon_{3}\right)$,$\left(F_{1},F_{2},\epsilon_{3}\right)$,$\left(F_{1},F_{2},\epsilon_{1},\epsilon_{3}\right)$
,$\left(F_{1},F_{2},\epsilon_{2},\epsilon_{3}\right),\left(F_{3},\epsilon_{2},\epsilon_{3}\right),\left(F_{1},F_{3},\epsilon_{3}\right)$,$\left(F_{2},F_{3},\epsilon_{3}\right)$
and $\left(F_{1},F_{2},F_{3},\epsilon_{3}\right)$. The MCMs as defined
above include all proper subsets, and not only those that have the
fewest number of terms. The simplest such MCMs, with one forcing and
one dissipation term, involve nonzero $\left(F_{1},\epsilon_{3}\right)$
and $\left(F_{2},\epsilon_{3}\right)$. In addition, there is nonzero
$\left(F_{3},\epsilon_{2},\epsilon_{3}\right)$, which is irreducible
to the other two MCMs owing to different placement of forcing. Thus
there are $3$ MCMs for subclass $2$: $\left(F_{1},\epsilon_{3}\right)$,
$\left(F_{2},\epsilon_{3}\right)$, and $\left(F_{3},\epsilon_{2},\epsilon_{3}\right)$,
for $p+q+r\neq0$.

Similar analysis for subclasses $3$ and $4$ identifies $11$ and
$8$ chaotic cases respectively, whose orbits are shown in Figures
3--4. For both of these subclasses, there are $3$ MCMs, with nonzero
$\left(F_{1},\epsilon_{3}\right)$, $\left(F_{2},\epsilon_{3}\right)$,
and $\left(F_{3},\epsilon_{3}\right)$. The corresponding time-series
of $x_{1}$, Poincar$\acute{\textrm{e}}$ sections, power spectra
of the stationary orbits, and evolution of LLE, are shown in SI Figures
11-14, and Figures 15-18, respectively for the two subclasses.

Table 1 below summarizes the chaotic cases and MCMs, for each of the
subclasses. It lists each of the chaotic cases for subclasses 1-4,
identified by the multiple lines of evidence (Section 2), and inspection
of these cases identifies the MCMs. While siting of forcing can vary
between subclasses, the most basic condition on dissipation is the
same and all chaotic cases must involve dissipation in the third (i.e.,
linear) equation through nonzero $\epsilon_{3}$.

\subsection*{With energy conservation constraint, i.e., $p+q+r=0$:}

Table 1 also lists the MCMs when the energy conservation constraint
is present, based on simulations with $q=-p$, since $r=0$. This
is based on a complete listing of cases with positive LLEs (SI Figures
22-24, 25-27, 28-30, and 31-33 for subclasses $1-4$ respectively),
for which we have used the aforementioned criteria to identify chaotic
cases and thereby establish MCMs. Table 1 below shows that generally
there are fewer MCMs when energy is conserved.

For subclass $1$, there is no effect of the energy conservation constraint,
with MCM $\left(F_{1},\epsilon_{3}\right)$ in either case.

In subclass $2$, only $\left(F_{1},\epsilon_{3}\right)$ and $\left(F_{2},\epsilon_{3}\right)$
are MCMs when energy is conserved.

For subclass $3$, $\left(F_{1},\epsilon_{3}\right)$ is an MCM while
$\left(F_{2},\epsilon_{3}\right)$ and $\left(F_{3},\epsilon_{3}\right)$
are not.

Similarly, in subclass $4$, $\left(F_{1},\epsilon_{3}\right)$ and
$\left(F_{2},\epsilon_{3}\right)$ are MCMs, while $\left(F_{3},\epsilon_{3}\right)$
is not.

If the gyrostat core conserves energy, forcing can be placed in fewer
ways for the model to admit chaos. These effects of the energy conservation
constraint on the MCMs are explained in the following section. When
energy is conserved in the gyrostat core, the MCMs are governed by
the arrangement of fixed points $\left(x_{1}^{*},x_{2}^{*},x_{3}^{*}\right)$
in the model with F\&D. For each subclass, the resulting conditions
for MCMs are found by inspecting the cubic equation for $x_{3}^{*}$,
whose coefficients are shown in the last column of Table 1. This is
because chaos for the energy conserving case requires a pair of fixed
points with opposite signs of $x_{3}^{*}$.

\pagebreak{}

\begin{landscape}

Table 1: Chaotic cases, minimal chaotic models (MCMs), and coefficients
of cubic equation describing $x_{3}^{*}$ for subclasses $1-4$.

\begin{tabular}{|c|c|c|c|c|}
\hline 
{\footnotesize{}Subclass} & {\footnotesize{}Chaotic cases $\left(p+q+r\neq0\right)$} & {\footnotesize{}MCM $\left(p+q+r\neq0\right)$} & {\footnotesize{}MCM $\left(p+q+r=0\right)$} & {\footnotesize{}Coefficients of cubic equation for $x_{3}^{*}$}\tabularnewline
\hline 
\hline 
\multirow{4}{*}{{\footnotesize{}$1$}} & \multirow{1}{*}{{\footnotesize{}$\left(F_{1},\epsilon_{3}\right)$, $\left(F_{1},\epsilon_{1},\epsilon_{3}\right)$,
$\left(F_{1},\epsilon_{2},\epsilon_{3}\right)$,}} & \multirow{4}{*}{{\footnotesize{}$\left(F_{1},\epsilon_{3}\right)$}} & \multirow{4}{*}{{\footnotesize{}$\left(F_{1},\epsilon_{3}\right)$}} & {\footnotesize{}$\gamma_{3}=\epsilon_{3}\frac{p}{a}$}\tabularnewline
\cline{2-2} \cline{5-5} 
 & {\footnotesize{}$\left(F_{1},F_{2},\epsilon_{3}\right)$, $\left(F_{1},F_{3},\epsilon_{3}\right)$,} &  &  & {\footnotesize{}$\gamma_{2}=-F_{3}\frac{p}{a}$,}\tabularnewline
\cline{2-2} \cline{5-5} 
 & {\footnotesize{}$\left(F_{1},F_{2},F_{3},\epsilon_{3}\right)$,} &  &  & {\footnotesize{}$\gamma_{1}=F_{1}-\epsilon_{1}\frac{a}{q}-\epsilon_{1}\epsilon_{2}\epsilon_{3}\frac{1}{aq}$,}\tabularnewline
\cline{2-2} \cline{5-5} 
 & {\footnotesize{}$\left(F_{1},F_{2},F_{3},\epsilon_{2},\epsilon_{3}\right)$.} &  &  & {\footnotesize{}$\gamma_{0}=\frac{F_{2}\epsilon_{1}}{q}+\frac{F_{3}\epsilon_{1}\epsilon_{2}}{aq}$.}\tabularnewline
\hline 
\multirow{4}{*}{{\footnotesize{}$2$}} & {\footnotesize{}$\left(F_{1},\epsilon_{3}\right)$, $\left(F_{1},\epsilon_{1},\epsilon_{3}\right)$,
$\left(F_{1},\epsilon_{2},\epsilon_{3}\right)$,} & \multirow{4}{*}{{\footnotesize{}$\left(F_{1},\epsilon_{3}\right),$$\left(F_{2},\epsilon_{3}\right)$,$\left(F_{3},\epsilon_{2},\epsilon_{3}\right)$}} & \multirow{4}{*}{{\footnotesize{}$\left(F_{1},\epsilon_{3}\right),$$\left(F_{2},\epsilon_{3}\right)$}} & {\footnotesize{}$\gamma_{3}=\epsilon_{3}\frac{p}{a}$,}\tabularnewline
\cline{2-2} \cline{5-5} 
 & {\footnotesize{}$\left(F_{2},\epsilon_{3}\right)$, $\left(F_{2},\epsilon_{1},\epsilon_{3}\right)$,$\left(F_{1},F_{2},\epsilon_{3}\right)$,} &  &  & {\footnotesize{}$\gamma_{2}=-F_{3}\frac{p}{a}+b\left(1+\frac{p}{q}\right)$,}\tabularnewline
\cline{2-2} \cline{5-5} 
 & {\footnotesize{}$\left(F_{1},F_{2},\epsilon_{1},\epsilon_{3}\right)$,$\left(F_{1},F_{2},\epsilon_{2},\epsilon_{3}\right),\left(F_{3},\epsilon_{2},\epsilon_{3}\right),$} &  &  & {\footnotesize{}$\gamma_{1}=F_{1}-F_{2}b\frac{p}{aq}-\epsilon_{1}\frac{a}{q}-\epsilon_{1}\epsilon_{2}\epsilon_{3}\frac{1}{aq}-\epsilon_{2}\frac{b^{2}}{aq}$,}\tabularnewline
\cline{2-2} \cline{5-5} 
 & {\footnotesize{}$\left(F_{1},F_{3},\epsilon_{3}\right)$, $\left(F_{2},F_{3},\epsilon_{3}\right)$,
$\left(F_{1},F_{2},F_{3},\epsilon_{3}\right)$.} &  &  & {\footnotesize{}$\gamma_{0}=\frac{F_{2}\epsilon_{1}}{q}+\frac{F_{3}\epsilon_{1}\epsilon_{2}}{aq}-F_{1}\epsilon_{2}\frac{b}{aq}$.}\tabularnewline
\hline 
\multirow{4}{*}{{\footnotesize{}$3$}} & {\footnotesize{}$\left(F_{1},\epsilon_{3}\right)$, $\left(F_{1},\epsilon_{2},\epsilon_{3}\right)$,
$\left(F_{2},\epsilon_{3}\right)$,} & \multirow{4}{*}{{\footnotesize{}$\left(F_{1},\epsilon_{3}\right)$, $\left(F_{2},\epsilon_{3}\right)$,
$\left(F_{3},\epsilon_{3}\right)$}} & \multirow{4}{*}{{\footnotesize{}$\left(F_{1},\epsilon_{3}\right)$}} & {\footnotesize{}$\gamma_{3}=\epsilon_{3}\frac{p}{a}$,}\tabularnewline
\cline{2-2} \cline{5-5} 
 & {\footnotesize{}$\left(F_{2},\epsilon_{2},\epsilon_{3}\right)$,$\left(F_{1},F_{2},\epsilon_{3}\right)$,} &  &  & {\footnotesize{}$\gamma_{2}=-F_{3}\frac{p}{a}+\epsilon_{3}\frac{c}{a}\left(\frac{p}{q}-1\right)$,}\tabularnewline
\cline{2-2} \cline{5-5} 
 & {\footnotesize{}$\left(F_{1},F_{2},\epsilon_{1},\epsilon_{3}\right)$,$\left(F_{3},\epsilon_{3}\right),\left(F_{1},F_{3},\epsilon_{3}\right)$,} &  &  & {\footnotesize{}$\gamma_{1}=F_{1}-F_{3}\frac{c}{a}\left(\frac{p}{q}-1\right)-\epsilon_{1}\frac{a}{q}-\epsilon_{1}\epsilon_{2}\epsilon_{3}\frac{1}{aq}-\epsilon_{3}\frac{c^{2}}{aq}$,}\tabularnewline
\cline{2-2} \cline{5-5} 
 & {\footnotesize{}$\left(F_{2},F_{3},\epsilon_{3}\right)$, $\left(F_{2},F_{3},\epsilon_{2},\epsilon_{3}\right)$,
$\left(F_{1},F_{2},F_{3},\epsilon_{3}\right)$.} &  &  & {\footnotesize{}$\gamma_{0}=F_{1}\frac{c}{q}+\frac{F_{2}\epsilon_{1}}{q}+F_{3}\left(\frac{c^{2}}{aq}+\frac{\epsilon_{1}\epsilon_{2}}{aq}\right)$.}\tabularnewline
\cline{2-5} \cline{3-5} \cline{4-5} \cline{5-5} 
\multicolumn{1}{c|}{\multirow{4}{*}{{\footnotesize{}$4$}}} & {\footnotesize{}$\left(F_{1},\epsilon_{3}\right)$, $\left(F_{2},\epsilon_{3}\right)$,
$\left(F_{1},F_{2},\epsilon_{3}\right)$,} & \multirow{4}{*}{{\footnotesize{}$\left(F_{1},\epsilon_{3}\right)$, $\left(F_{2},\epsilon_{3}\right)$,
$\left(F_{3},\epsilon_{3}\right)$}} & \multirow{4}{*}{{\footnotesize{}$\left(F_{1},\epsilon_{3}\right)$, $\left(F_{2},\epsilon_{3}\right)$}} & {\footnotesize{}$\gamma_{3}=\epsilon_{3}\frac{p}{a}$}\tabularnewline
\cline{2-2} \cline{5-5} 
 & {\footnotesize{}$\left(F_{1},F_{2},\epsilon_{2},\epsilon_{3}\right)$,$\left(F_{3},\epsilon_{3}\right),\left(F_{1},F_{3},\epsilon_{3}\right)$,} &  &  & {\footnotesize{}$\gamma_{2}=-F_{3}\frac{p}{a}+\epsilon_{3}\frac{c}{a}\left(\frac{p}{q}-1\right)+b\left(1+\frac{px_{1}^{*}}{a}\right)$,}\tabularnewline
\cline{2-2} \cline{5-5} 
 & {\footnotesize{}$\left(F_{2},F_{3},\epsilon_{3}\right)$, $\left(F_{1},F_{2},F_{3},\epsilon_{3}\right)$.} &  &  & {\footnotesize{}$\gamma_{1}=F_{1}-F_{3}\frac{c}{a}\left(\frac{p}{q}-1\right)-\epsilon_{1}\frac{a}{q}-\epsilon_{1}\epsilon_{2}\epsilon_{3}\frac{1}{aq}-\epsilon_{3}\frac{c^{2}}{aq}+\frac{bc}{q}+\frac{bcx_{1}^{*}}{a}\left(\frac{p}{q}-1\right)$,}\tabularnewline
\cline{2-2} \cline{5-5} 
 &  &  &  & {\footnotesize{}$\gamma_{0}=F_{1}\frac{c}{q}+\frac{F_{2}\epsilon_{1}}{q}+F_{3}\left(\frac{c^{2}}{aq}+\frac{\epsilon_{1}\epsilon_{2}}{aq}\right)-\frac{bx_{1}^{*}\left(c^{2}+\epsilon_{1}\epsilon_{2}\right)}{aq}$.}\tabularnewline
\cline{2-5} \cline{3-5} \cline{4-5} \cline{5-5} 
\end{tabular}

\end{landscape}

\pagebreak{}

\section{Accounting for minimal chaotic models}

\subsection{Subclass 1 with $r=0,b=c=0$}

This subclass of the Volterra gyrostat has evolution
\begin{align}
x_{1}' & =px_{2}x_{3}\nonumber \\
x_{2}' & =qx_{3}x_{1}-ax_{3}.\nonumber \\
x_{3}' & =ax_{2}\label{eq:p5}
\end{align}
and serves as the conservative core of many important LOMs (\citet{Gluhovsky1999}),
including that of \citet{Lorenz1963}. It has two invariants irrespective
of whether $p+q=0$ (\citet{Seshadri2023}), and rotational symmetry
$R_{x_{1}}\left(\pi\right)$ about the axis of $x_{1}$, i.e. the
equations are preserved under the transformation $\left(x_{1},x_{2},x_{3}\right)\rightarrow\left(x_{1},-x_{2},-x_{3}\right)$.
All chaotic cases have nonzero $\left(F_{1},\epsilon_{3}\right)$
and the MCM
\begin{align}
x_{1}' & =px_{2}x_{3}+F_{1}\nonumber \\
x_{2}' & =qx_{3}x_{1}-ax_{3}.\nonumber \\
x_{3}' & =ax_{2}-\epsilon_{3}x_{3}\label{eq:p6}
\end{align}
maintains the symmetry $R_{x_{1}}\left(\pi\right)$. For this last
system there are two distinct fixed points given by $\left(a/q,x_{2}^{*},x_{3}^{*}\right)$
and $\left(a/q,-x_{2}^{*},-x_{3}^{*}\right)$ with $x_{2}^{*2}=-\frac{\epsilon_{3}F_{1}}{ap}$
and $x_{3}^{*2}=-\frac{aF_{1}}{p\epsilon_{3}}$, which are real if
$F_{1}$ has opposite sign from $ap$ $\left(\epsilon_{3}>0\right)$.
The Jacobian evaluated at $\left(a/q,x_{2}^{*},x_{3}^{*}\right)$
becomes 
\begin{equation}
\mathrm{D}_{\mathrm{f}}=\left[\begin{array}{ccc}
0 & px_{3}^{*} & px_{2}^{*}\\
qx_{3}^{*} & 0 & 0\\
0 & a & -\epsilon_{3}
\end{array}\right]\label{eq:p7}
\end{equation}
having characteristic polynomial 
\begin{equation}
\lambda^{3}+\epsilon_{3}\lambda^{2}-pqx_{3}^{*2}\lambda-2\epsilon_{3}pqx_{3}^{*2}=0,\label{eq:p8}
\end{equation}
where we have used, from the last component of Eq. (\ref{eq:p6}),
$x_{2}^{*}=\left(\epsilon_{3}/a\right)x_{3}^{*}$. From the symmetry
of the equations, the Jacobian evaluated at $\left(a/q,-x_{2}^{*},-x_{3}^{*}\right)$
also has the same characteristic polynomial, and thus the above fixed
points make a pair with identical stability. Moreover the discriminant
of the characteristic polynomial in Eq. (\ref{eq:p8}) is negative\footnote{For a general cubic given by $\beta_{3}x^{3}+\beta_{2}x^{2}+\beta_{1}x+\beta_{0}=0,$the
discriminant is defined as $\Delta=18\beta_{3}\beta_{2}\beta_{1}\beta_{0}-4\beta_{2}^{3}\beta_{0}+\beta_{2}^{2}\beta_{1}^{2}-4\beta_{3}\beta_{1}^{3}-27\beta_{3}^{2}\beta_{0}^{2}$
and its sign determines the number of real and complex roots.} 
\begin{equation}
\Delta=pqx_{3}^{*2}\left(8\epsilon_{3}^{4}-71\epsilon_{3}^{2}pqx_{3}^{*2}+4p^{2}q^{2}x_{3}^{*4}\right)<0\label{eq:p9}
\end{equation}
in case $pq<0$, since $x_{3}^{*2}>0$. The MCM plotted in Figure
1 has parameters $a=-0.94$, $p=0.19$, and $q=-0.39$, so that indeed
$pq<0$. Therefore there are two complex eigenvalues at each fixed
point, with a Hopf bifurcation occurring when the real part crosses
the imaginary axis. With product of all eigenvalues being negative
there is one stable direction, but with large $F_{1}$ the real part
of the complex conjugate pair becomes positive and the fixed points
repel nearby trajectories.

These properties are not maintained in case a single forcing and dissipation
term occur elsewhere. Table 2 lists the $8$ alternate ways in which
a single forcing and a single dissipation term can be placed. None
of these can admit chaos. Each of these yields a different arrangement
of fixed points from the above: ranging from no fixed points to an
entire line. Where a pair of fixed points exists, they do not possess
the above symmetry. The MCM $\left(F_{1},\epsilon_{3}\right)$ is
not alone in maintaining the symmetry of the gyrostat core, and other
non-chaotic cases in Table 2, such as $\left(F_{1},\epsilon_{1}\right)$
and $\left(F_{1},\epsilon_{2}\right)$, maintain it as well. The MCM
$\left(F_{1},\epsilon_{3}\right)$ has the unique attribute of maintaining
symmetry of the equations while giving rise to a pair of distinct
and symmetrically placed fixed points with opposite signs of $x_{3}^{*}$.
Other chaotic cases listed in Table 1 do not always maintain this
symmetry. As shown below, the necessary condition for appearance of
chaos in this subclass is a pair of fixed points arranged on opposite
sides of the $x_{1}-x_{2}$ plane. The MCM $\left(F_{1},\epsilon_{3}\right)$
also appears when energy is conserved in the gyrostat core (SI Figs.
22-24; Table 1).

\pagebreak{}

\begin{landscape}

Table 2: Fixed points for various other placements of one F\&D term
(besides the MCM) in Subclass 1.

\begin{tabular}{|c|c|c|c|}
\hline 
S. No & Model & Equations & Fixed Points\tabularnewline
\hline 
$1$ & $F_{1},\epsilon_{1}$ & $x_{1}'=px_{2}x_{3}+F_{1}-\epsilon_{1}x_{1}\textrm{, }x_{2}'=qx_{3}x_{1}-ax_{3}\textrm{, }x_{3}'=ax_{2}$ & $\left(\frac{F_{1}}{\epsilon_{1}},0,0\right)$\tabularnewline
\hline 
$2$ & $F_{1},\epsilon_{2}$ & $x_{1}'=px_{2}x_{3}+F_{1}\textrm{, }x_{2}'=qx_{3}x_{1}-ax_{3}-\epsilon_{2}x_{2}\textrm{, }x_{3}'=ax_{2}$ & None\tabularnewline
\hline 
$3$ & $F_{2},\epsilon_{1}$ & $x_{1}'=px_{2}x_{3}-\epsilon_{1}x_{1}\textrm{, }x_{2}'=qx_{3}x_{1}-ax_{3}+F_{2}\textrm{, }x_{3}'=ax_{2}$ & $\left(0,0,\frac{F_{2}}{a}\right)$\tabularnewline
\hline 
$4$ & $F_{2},\epsilon_{2}$ & $x_{1}'=px_{2}x_{3}\textrm{, }x_{2}'=qx_{3}x_{1}-ax_{3}+F_{2}-\epsilon_{2}x_{2}\textrm{, }x_{3}'=ax_{2}$ & $\left(x_{1}^{*},0,-\frac{F_{2}}{qx_{1}^{*}-a}\right)$\tabularnewline
\hline 
$5$ & $F_{2},\epsilon_{3}$ & $x_{1}'=px_{2}x_{3}\textrm{, }x_{2}'=qx_{3}x_{1}-ax_{3}+F_{2}\textrm{, }x_{3}'=ax_{2}-\epsilon_{3}x_{3}$ & None\tabularnewline
\hline 
$6$ & $F_{3},\epsilon_{1}$ & $x_{1}'=px_{2}x_{3}-\epsilon_{1}x_{1}\textrm{, }x_{2}'=qx_{3}x_{1}-ax_{3}\textrm{, }x_{3}'=ax_{2}+F_{3}$ & $\left(0,-\frac{F_{3}}{a},0\right)$ and $\left(\frac{a}{q},-\frac{F_{3}}{a},-\frac{\epsilon_{1}a^{2}}{pqF_{3}}\right)$\tabularnewline
\hline 
$7$ & $F_{3},\epsilon_{2}$ & $x_{1}'=px_{2}x_{3}\textrm{, }x_{2}'=qx_{3}x_{1}-ax_{3}-\epsilon_{2}x_{2}\textrm{, }x_{3}'=ax_{2}+F_{3}$ & None\tabularnewline
\hline 
$8$ & $F_{3},\epsilon_{3}$ & $x_{1}'=px_{2}x_{3}\textrm{, }x_{2}'=qx_{3}x_{1}-ax_{3}\textrm{, }x_{3}'=ax_{2}+F_{3}-\epsilon_{3}x_{3}$ & $\left(\frac{a}{q},0,\frac{F_{3}}{\epsilon_{3}}\right)$ and $\left(x_{1}^{*},-\frac{F_{3}}{a},0\right)$\tabularnewline
\hline 
\end{tabular}

\end{landscape}

\pagebreak{}

More generally, this subclass with F\&D
\begin{align}
x_{1}' & =px_{2}x_{3}-\epsilon_{1}x_{1}+F_{1}\nonumber \\
x_{2}' & =qx_{3}x_{1}-ax_{3}-\epsilon_{2}x_{2}+F_{2}\nonumber \\
x_{3}' & =ax_{2}-\epsilon_{3}x_{3}+F_{3}\label{eq:p10}
\end{align}
has fixed points $\left(x_{1}^{*},x_{2}^{*},x_{3}^{*}\right)$ as
follows: from the third equation, $x_{2}^{*}=-\left(F_{3}-\epsilon_{3}x_{3}^{*}\right)/a$
and from the second equation $qx_{1}^{*}x_{3}^{*}=ax_{3}^{*}+\epsilon_{2}x_{2}^{*}-F_{2}$.
Together these equations must obey consistency 
\begin{equation}
\left(qx_{1}^{*}-\frac{\epsilon_{2}\epsilon_{3}}{a}-a\right)x_{3}^{*}=-\frac{F_{3}\epsilon_{2}}{a}-F_{2}\label{eq:p11}
\end{equation}
so that when $x_{3}^{*}=0$ the right hand side of Eq. (\ref{eq:p11})
must also vanish. A second consistency condition is furnished by the
first equation 
\begin{equation}
px_{2}^{*}x_{3}^{*}=-F_{1}+\epsilon_{1}x_{1}^{*}\label{eq:p12}
\end{equation}
so that zero $x_{3}^{*}$ in conjunction with nonzero $F_{1}$ also
entails nonzero $\epsilon_{1}$. These consistency conditions are
met by the cases illustrated in Table 1. Using these relations to
eliminate $x_{1}^{*},x_{2}^{*}$ from the first equation we obtain
a cubic in $x_{3}^{*}$ 
\begin{align}
\gamma_{3}x_{3}^{*3}+\gamma_{2}x_{3}^{*2}+\gamma_{1}x_{3}^{*}+\gamma_{0} & =0\label{eq:p13}
\end{align}
where 
\begin{align}
\gamma_{3} & =\epsilon_{3}\frac{p}{a},\nonumber \\
\gamma_{2} & =-F_{3}\frac{p}{a},\nonumber \\
\gamma_{1} & =F_{1}-\epsilon_{1}\frac{a}{q}-\epsilon_{1}\epsilon_{2}\epsilon_{3}\frac{1}{aq},\nonumber \\
\gamma_{0} & =\frac{F_{2}\epsilon_{1}}{q}+\frac{F_{3}\epsilon_{1}\epsilon_{2}}{aq}.\label{eq:p14}
\end{align}
The number and arrangement of fixed points depends on the number of
real roots of Eq. (\ref{eq:p13}), subject to the two consistency
conditions. Recall that the MCM has nonzero $\left(F_{1},\epsilon_{3}\right)$.
The resulting expression becomes
\begin{equation}
\epsilon_{3}\frac{p}{a}x_{3}^{*3}+F_{1}x_{3}^{*}=0
\end{equation}
yielding two roots satisfying $x_{3}^{*2}=-\frac{aF_{1}}{p\epsilon_{3}}$
as found above.\footnote{The third root $x_{3}^{*}=0$ does not meet consistency condition
in Eq. (\ref{eq:p12}).} This is a simple model with F\&D yielding two fixed points with positive
and negative $x_{3}^{*}$. In fact, the MCM $\left(F_{1},\epsilon_{3}\right)$
provides the simplest model giving a pair of fixed points, with opposite
signs of $x_{3}^{*}$, which can undergo a Hopf bifurcation.\footnote{To illustrate why this cannot happen with another case having two
fixed points, such as $\left(F_{3},\epsilon_{3}\right)$: we obtain
$\epsilon_{3}\frac{p}{a}x_{3}^{*3}-F_{3}\frac{p}{a}x_{3}^{*2}=0$,
giving a fixed point $x_{3}^{*}=0$ and $x_{3}^{*}=F_{3}/\epsilon_{3}$.
What precludes it becoming chaotic? Owing to the same placement of
dissipation the model has the same underlying Jacobian and resulting
characteristic polynomial as Eq. (\ref{eq:p8}), and consequently
the same discriminant as Eq. (\ref{eq:p9}). At $x_{3}^{*}=0$ clearly
$\Delta=0$ and the equation has a double eigenvalue, precluding complex
roots. Thus a Hopf bifurcation cannot arise in this case. A similar
calculation precludes Hopf bifurcation in the pair of fixed points
with $\left(F_{3},\epsilon_{1}\right)$ in Table 1.}

Such fixed points play an important role in the Lorenz model (\citet{Sparrow1982}).
On the route to chaos, as the forcing is increased, following the
Hopf bifurcation, trajectories spiral around these unstable fixed
points at an increasing distance (\citet{Kaplan1979}). For subclass
$1$, the existence of this pair is central to the MCM.

There is an important difference with the Lorenz model, which must
be pointed out: the absence of a third fixed point, a saddle, for
the MCM precludes homoclinic orbits that are important in the Lorenz
model's dynamics (\citet{Sparrow1982}). Despite this, chaos arises
on account of the pair of fixed points, with positive/negative $x_{3}^{*}$,
yielding opposing spirals when projected onto the $x_{1}-x_{2}$ plane.\footnote{The role of the pair of fixed points with opposite signs of $x_{3}^{*}$
in creating opposing spirals when projected onto $x_{1}-x_{2}$ is
clear from the Jacobian in Eq. (\ref{eq:p7}).}

\subsection{Subclass 2 with $r=0,c=0$}

Here the conservative core 
\begin{align}
x_{1}' & =px_{2}x_{3}+bx_{3}\nonumber \\
x_{2}' & =qx_{3}x_{1}-ax_{3}\nonumber \\
x_{3}' & =ax_{2}-bx_{1}\label{eq:p16}
\end{align}
has a single invariant (\citet{Seshadri2023}) in general with $p+q\neq0$,
and, compared to subclass $1$, $b\neq0$ breaks the symmetry $R_{x_{1}}\left(\pi\right)$.
The MCMs $\left(F_{1},\epsilon_{3}\right)$ and $\left(F_{2},\epsilon_{3}\right)$,
which both appear even with $p+q+r=0$ (Table 1), are analogous, so
we consider $\left(F_{1},\epsilon_{3}\right)$
\begin{align}
x_{1}' & =px_{2}x_{3}+bx_{3}+F_{1}\nonumber \\
x_{2}' & =qx_{3}x_{1}-ax_{3}\nonumber \\
x_{3}' & =ax_{2}-bx_{1}-\epsilon_{3}x_{3}\label{eq:p17}
\end{align}
whose fixed points are given by $\left(\frac{a}{q},\frac{b}{q}+\frac{\epsilon_{3}}{a}x_{3}^{*},x_{3}^{*}\right)$
where $x_{3}^{*}$ solves the quadratic equation $\epsilon_{3}pqx_{3}^{*2}+ab\left(p+q\right)x_{3}^{*}+aqF_{1}=0$.
The Jacobian evaluated at the fixed point
\begin{equation}
\mathrm{D}_{\mathrm{f}}=\left[\begin{array}{ccc}
0 & px_{3}^{*} & px_{2}^{*}+b\\
qx_{3}^{*} & 0 & 0\\
-b & a & -\epsilon_{3}
\end{array}\right]\label{eq:p18}
\end{equation}
has eigenvalues solving characteristic polynomial
\begin{equation}
\lambda^{3}+\epsilon_{3}\lambda^{2}-\left(pqx_{3}^{*2}-b\right)\lambda-\epsilon_{3}pqx_{3}^{*2}+aqF_{1}=0,\label{eq:p22}
\end{equation}

which depends on $x_{3}^{*2}$. For $p+q\neq0$ the two fixed points
do not have identical stability. As with subclass $1$, for $F_{1}$
sufficiently large the complex eigenvalues have positive real part
so this pair of fixed points repels the flow in the neighborhoods.
The key condition for these MCMs is, again, the pair of fixed points
with opposite signs of $x_{3}^{*}$.

In contrast, the other MCM $\left(F_{3},\epsilon_{2},\epsilon_{3}\right)$
\begin{align}
x_{1}' & =px_{2}x_{3}+bx_{3}\nonumber \\
x_{2}' & =qx_{3}x_{1}-ax_{3}-\epsilon_{2}x_{2}\nonumber \\
x_{3}' & =ax_{2}-bx_{1}-\epsilon_{3}x_{3}+F_{3}\label{eq:p20}
\end{align}
does not appear in the presence of energy conservation, and cannot
be rooted in the aforementioned property of fixed points, which does
not hold (as shown below).

Therefore, with the energy conservation constraint, let us examine
the $7$ alternate placements of a single forcing and dissipation
term in Table 3. Most of the alternate placements yield only a single
fixed point, whereas those with two fixed points have one of them
lying on the $x_{1}-x_{2}$ plane. This does not favor the pair of
opposing spirals described above. Only the MCMs $\left(F_{1},\epsilon_{3}\right)$
and $\left(F_{2},\epsilon_{3}\right)$ lead to a pair of fixed points
with opposite signs of $x_{3}^{*}$, and these are the ones leading
to the pair of opposing spirals (see Eq. (\ref{eq:p18})). It is therefore
not surprising that these are the MCMs obtained with $p+q+r=0$ (SI
Figs. 25-27; Table 1). In summary, with the energy conservation constraint
in the gyrostat core, the requirement of a pair of fixed points with
opposite $x_{3}^{*}$ circumscribes the MCMs.

\pagebreak{}

\begin{landscape}

Table 3: Fixed points for various other placements of one forcing
and dissipation term in Subclass 2. Here we have taken $p+q+r=0$.

\begin{tabular}{|c|c|c|c|}
\hline 
S. No & Model & Equations & Fixed Points\tabularnewline
\hline 
$1$ & $F_{1},\epsilon_{1}$ & $x_{1}'=px_{2}x_{3}+bx_{3}+F_{1}-\epsilon_{1}x_{1}\textrm{, }x_{2}'=qx_{3}x_{1}-ax_{3}\textrm{, }x_{3}'=ax_{2}-bx_{1}$ & $\left(\frac{F_{1}}{\epsilon_{1}},\frac{bF_{1}}{a\epsilon_{1}},0\right)$\tabularnewline
\hline 
$2$ & $F_{1},\epsilon_{2}$ & $x_{1}'=px_{2}x_{3}+bx_{3}+F_{1}\textrm{, }x_{2}'=qx_{3}x_{1}-ax_{3}-\epsilon_{2}x_{2}\textrm{, }x_{3}'=ax_{2}-bx_{1}$ & $\left(\frac{a}{b}x_{2}^{*},x_{2}^{*}=-\frac{bx_{3}^{*}+F_{1}}{px_{3}^{*}},x_{3}^{*}=\frac{F_{1}\epsilon_{2}b}{F_{1}aq-\epsilon_{2}b}\right)$\tabularnewline
\hline 
$3$ & $F_{2},\epsilon_{1}$ & $x_{1}'=px_{2}x_{3}+bx_{3}-\epsilon_{1}x_{1}\textrm{, }x_{2}'=qx_{3}x_{1}-ax_{3}+F_{2}\textrm{, }x_{3}'=ax_{2}-bx_{1}$ & $\left(\frac{a}{b}x_{2}^{*},x_{2}^{*}=-\frac{bx_{3}^{*}}{px_{3}^{*}-\epsilon_{1}\frac{a}{b}},x_{3}^{*}=\frac{F_{2}\epsilon_{1}a}{F_{2}b-\epsilon_{1}a^{2}}\right)$\tabularnewline
\hline 
$4$ & $F_{2},\epsilon_{2}$ & $x_{1}'=px_{2}x_{3}+bx_{3}\textrm{, }x_{2}'=qx_{3}x_{1}-ax_{3}+F_{2}-\epsilon_{2}x_{2}\textrm{, }x_{3}'=ax_{2}-bx_{1}$ & $\left(\frac{a}{b}\frac{F_{2}}{\epsilon_{2}},\frac{F_{2}}{\epsilon_{2}},0\right)$\tabularnewline
\hline 
$5$ & $F_{3},\epsilon_{1}$ & $x_{1}'=px_{2}x_{3}+bx_{3}-\epsilon_{1}x_{1}\textrm{, }x_{2}'=qx_{3}x_{1}-ax_{3}\textrm{, }x_{3}'=ax_{2}-bx_{1}+F_{3}$ & $\left(0,-\frac{F_{3}}{a},0\right)$ and $\left(\frac{a}{q},-\frac{F_{3}-\frac{ba}{q}}{a},-\frac{\epsilon_{1}a^{2}}{pqF_{3}}\right)$\tabularnewline
\hline 
$6$ & $F_{3},\epsilon_{2}$ & $x_{1}'=px_{2}x_{3}+bx_{3}\textrm{, }x_{2}'=qx_{3}x_{1}-ax_{3}-\epsilon_{2}x_{2}\textrm{, }x_{3}'=ax_{2}-bx_{1}+F_{3}$ & $\left(\frac{F_{3}}{b},0,0\right)$ and $\left(\frac{-\frac{ab}{p}+F_{3}}{b},-\frac{b}{p},-\frac{\epsilon_{2}b^{2}}{pqF_{3}}\right)$\tabularnewline
\hline 
$7$ & $F_{3},\epsilon_{3}$ & $x_{1}'=px_{2}x_{3}+bx_{3}\textrm{, }x_{2}'=qx_{3}x_{1}-ax_{3}\textrm{, }x_{3}'=ax_{2}-bx_{1}+F_{3}-\epsilon_{3}x_{3}$ & $\left(x_{1}^{*},-\frac{F_{3}-bx_{1}^{*}}{a},0\right)$ and $\left(\frac{a}{q},-\frac{b}{p},\frac{F_{3}}{\epsilon_{3}}\right)$\tabularnewline
\hline 
\end{tabular}

\end{landscape}

\pagebreak{}

Here the possible solutions for $x_{3}^{*}$ for general placement
of forcing and dissipation follows cubic
\begin{align}
\gamma_{3}x_{3}^{*3}+\gamma_{2}x_{3}^{*2}+\gamma_{1}x_{3}^{*}+\gamma_{0} & =0\label{eq:p21}
\end{align}
where 
\begin{align}
\gamma_{3} & =\epsilon_{3}\frac{p}{a},\nonumber \\
\gamma_{2} & =-F_{3}\frac{p}{a}+b\left(1+\frac{p}{q}\right),\nonumber \\
\gamma_{1} & =F_{1}-F_{2}b\frac{p}{aq}-\epsilon_{1}\frac{a}{q}-\epsilon_{1}\epsilon_{2}\epsilon_{3}\frac{1}{aq}-\epsilon_{2}\frac{b^{2}}{aq},\nonumber \\
\gamma_{0} & =\frac{F_{2}\epsilon_{1}}{q}+\frac{F_{3}\epsilon_{1}\epsilon_{2}}{aq}-F_{1}\epsilon_{2}\frac{b}{aq},
\end{align}
which reduces to Eq. (\ref{eq:p14}) for $b=0$.

The MCMs for the energy conserving core generally describe the simplest
ways to obtain a pair of fixed points with opposite signs of $x_{3}^{*}$,
and nonzero $\epsilon_{3}$ along with suitable placement of forcing
supports this. Nonzero $\left(F_{1},\epsilon_{3}\right)$ and $\left(F_{2},\epsilon_{3}\right)$
both yield $\gamma_{3}x_{3}^{*3}+\gamma_{1}x_{3}^{*}=0$ for the condition
$p+q=0$, giving a pair of fixed points having the aforementioned
property if $\gamma_{1}\gamma_{3}<0$.

The other MCM $\left(F_{3},\epsilon_{2},\epsilon_{3}\right)$ gives
$\gamma_{3}x_{3}^{*3}+\gamma_{2}x_{3}^{*2}+\gamma_{1}x_{3}^{*}=0$:
it does not fall into the same pattern as the above cases,\footnote{Here, it is easily seen that $\gamma_{1}$ and $\gamma_{3}$ are of
the same sign, precluding roots of opposite sign.} as nonzero $F_{3}$ precludes a second fixed point across the plane.
It is notable that this case does not occur with $p+q+r=0$ (SI Figs.
25-27; Table 1).

If only two successive coefficients are nonzero, no matter their degree,
this would give at most one non-trivial $x_{3}^{*}$. This occurs
with $\left(F_{3},\epsilon_{3}\right)$, giving $\gamma_{3}x_{3}^{*3}+\gamma_{2}x_{2}^{*2}=0$.
Other possibilities are also precluded, for example if the constant
and quadratic coefficients have the same sign.\footnote{This occurs with nonzero $\text{\ensuremath{\left(F_{3},\epsilon_{1},\epsilon_{2}\right)}}$,
which gives $-F_{3}\frac{p}{a}x_{3}^{*2}-\left(\epsilon_{1}\frac{a}{q}+\epsilon_{2}\frac{b^{2}}{aq}\right)+\frac{F_{3}\epsilon_{1}\epsilon_{2}}{aq}=0$.
In order for the roots to have opposite sign, the constant and quadratic
coefficients must be of opposite sign, which is not possible since
$pq<0$. The case $\text{\ensuremath{\left(F_{3},F_{2},\epsilon_{1}\right)}}$
leads to the same difficulty, as the roots are given by $-F_{3}\frac{p}{a}x_{3}^{*2}-\left(F_{2}b\frac{p}{aq}+\epsilon_{1}\frac{a}{q}\right)+\frac{F_{2}\epsilon_{1}}{q}=0$,
with quadratic and constant coefficients of the same sign (from our
premise of $F_{3}=F_{2}$). Since the first two equations of this
subclass have the same form, we can also rule out chaos in $\text{\ensuremath{\left(F_{3},F_{1},\epsilon_{2}\right)}}$.} This rules out other fixed point arrangements besides the MCMs, for
an energy conserving core.

In summary, we have shown that the pair of fixed points with opposite
signs of $x_{3}^{*}$ cannot be achieved with a quadratic equation
for the roots, and nonzero $\gamma_{3}=\epsilon_{3}\frac{p}{a}$ is
required. This accounts for the presence of dissipation in the linear
equation, in all chaotic models. The cases $\left(F_{1},\epsilon_{3}\right)$
and $\left(F_{2},\epsilon_{3}\right)$ each yield pairs of fixed points
with opposite signs of $x_{3}^{*}$, and the corresponding dynamics
is obtained not only for $p+q=0$, but also for the core without energy
conservation. In contrast, the MCM $\left(F_{3},\epsilon_{2},\epsilon_{3}\right)$,
which appears only when the gyrostat core does not conserve energy,
cannot be tied to increasing repulsion of orbits by such a fixed point
pair, but is tied to the presence of only one quadratic invariant
when $p+q\neq0$ in subclass $2$.

\subsection{Subclass 3 with $r=0,b=0$ and Subclass 4 with $r=0$}

For subclass $3$ the equations 
\begin{align}
x_{1}' & =px_{2}x_{3}-cx_{2}-\epsilon_{1}x_{1}+F_{1}\nonumber \\
x_{2}' & =qx_{3}x_{1}-ax_{3}+cx_{1}-\epsilon_{2}x_{2}+F_{2}\nonumber \\
x_{3}' & =ax_{2}-\epsilon_{3}x_{3}+F_{3}\label{eq:p23}
\end{align}
give fixed points $x_{2}^{*}=-\left(F_{3}-\epsilon_{3}x_{3}^{*}\right)/a$,
and $x_{1}^{*}=\left(ax_{3}^{*}+\epsilon_{2}x_{2}^{*}-F_{2}\right)/\left(qx_{3}^{*}+c\right)$,
resulting in a cubic equation $\gamma_{3}x_{3}^{*3}+\gamma_{2}x_{3}^{*2}+\gamma_{1}x_{3}^{*}+\gamma_{0}=0$
as before, with coefficients
\begin{align}
\gamma_{3} & =\epsilon_{3}\frac{p}{a},\nonumber \\
\gamma_{2} & =-F_{3}\frac{p}{a}+\epsilon_{3}\frac{c}{a}\left(\frac{p}{q}-1\right),\nonumber \\
\gamma_{1} & =F_{1}-F_{3}\frac{c}{a}\left(\frac{p}{q}-1\right)-\epsilon_{1}\frac{a}{q}-\epsilon_{1}\epsilon_{2}\epsilon_{3}\frac{1}{aq}-\epsilon_{3}\frac{c^{2}}{aq},\nonumber \\
\gamma_{0} & =F_{1}\frac{c}{q}+\frac{F_{2}\epsilon_{1}}{q}+F_{3}\left(\frac{c^{2}}{aq}+\frac{\epsilon_{1}\epsilon_{2}}{aq}\right).
\end{align}
Here too, chaos requires nonzero $\gamma_{3}=\epsilon_{3}\frac{p}{a}$.
For the energy-conserving core, a pair of real solutions with positive/negative
$x_{3}^{*}$ cannot be realized with a quadratic equation alone.\footnote{For e.g., nonzero $\left(F_{3},\epsilon_{1}\right)$, $\left(F_{3},\epsilon_{2}\right),\left(F_{3},\epsilon_{1},\epsilon_{2}\right)$,
etc., give quadratic and constant coefficients of the same sign, and
no real solutions.}

In the presence of forcing, the minimal chaotic case $\left(F_{1},\epsilon_{3}\right)$
gives a cubic equation with all coefficients being nonzero, with many
settings of the parameters allowing the desired configuration of the
fixed-point pair.

In case of $\left(F_{2},\epsilon_{3}\right)$ we obtain $\gamma_{3}x_{3}^{*3}+\gamma_{2}x_{3}^{*2}+\gamma_{1}x_{3}^{*}=0$,
with $\gamma_{1},\gamma_{3}$ of same sign. Thus the necessary configuration
of nontrivial fixed points is not available, indicating other processes
at work. It is thus hardly surprising that this case is not a MCM
in case $p+q+r=0$ (SI Figs. 28-30; Table 1). The same goes for $\left(F_{3},\epsilon_{3}\right)$,
which requires nonconservation of energy in the gyrostat core to give
rise to F\&D chaos.

Finally we consider Subclass $4$, for which the F\&D equations
\begin{align}
x_{1}' & =px_{2}x_{3}+bx_{3}-cx_{2}-\epsilon_{1}x_{1}+F_{1}\nonumber \\
x_{2}' & =qx_{3}x_{1}-ax_{3}+cx_{1}-\epsilon_{2}x_{2}+F_{2}\nonumber \\
x_{3}' & =ax_{2}-bx_{1}-\epsilon_{3}x_{3}+F_{3}\label{eq:p25}
\end{align}
have fixed points $x_{2}^{*}=-\left(F_{3}-bx_{1}^{*}-\epsilon_{3}x_{3}^{*}\right)/a$,
and $x_{1}^{*}=\left(ax_{3}^{*}+\epsilon_{2}x_{2}^{*}-F_{2}\right)/\left(qx_{3}^{*}+c\right)$,
resulting in cubic $\gamma_{3}x_{3}^{*3}+\gamma_{2}x_{3}^{*2}+\gamma_{1}x_{3}^{*}+\gamma_{0}=0$
having coefficients
\begin{align}
\gamma_{3} & =\epsilon_{3}\frac{p}{a},\nonumber \\
\gamma_{2} & =-F_{3}\frac{p}{a}+\epsilon_{3}\frac{c}{a}\left(\frac{p}{q}-1\right)+b\left(1+\frac{px_{1}^{*}}{a}\right),\nonumber \\
\gamma_{1} & =F_{1}-F_{3}\frac{c}{a}\left(\frac{p}{q}-1\right)-\epsilon_{1}\frac{a}{q}-\epsilon_{1}\epsilon_{2}\epsilon_{3}\frac{1}{aq}-\epsilon_{3}\frac{c^{2}}{aq}+\frac{bc}{q}+\frac{bcx_{1}^{*}}{a}\left(\frac{p}{q}-1\right)\nonumber \\
\gamma_{0} & =F_{1}\frac{c}{q}+\frac{F_{2}\epsilon_{1}}{q}+F_{3}\left(\frac{c^{2}}{aq}+\frac{\epsilon_{1}\epsilon_{2}}{aq}\right)-\frac{bx_{1}^{*}\left(c^{2}+\epsilon_{1}\epsilon_{2}\right)}{aq},\label{eq:p26}
\end{align}
which closely parallels Subclass 3, with additional contributions
from nonzero $b$. Even though this is cubic the structure is really
different, since it must be solved simultaneously with the other equations
for $x_{1}^{*}$ and $x_{2}^{*}$, and yet the close resemblance points
to how the same MCMs are attained (Table 1), once $\epsilon_{3}$
is nonzero. There are subtle differences, however, which make not
only $\left(F_{1},\epsilon_{3}\right)$, but also $\left(F_{2},\epsilon_{3}\right)$
as MCMs, in the presence of an energy conserving core (SI Figs. 31-33;
Table 1). The inclusion of $\left(F_{2},\epsilon_{3}\right)$ can
be expected from the similarity of the equations for $dx_{1}/dt$
and $dx_{2}/dt$.

\subsection{Common features of MCMs across subclasses}

\subsubsection*{Role of fixed point pair with positive/negative $x_{3}^{*}$ for
the energy-conserving gyrostat core}

For subclass $1$, Figures 5-6 (and SI Figs. 19-20) illustrate the
orbits for the models $\left(F_{1},\epsilon_{3}\right)$, and those
for $\left(F_{1},\epsilon_{2},\epsilon_{3}\right)$. The latter closely
resembles the model of \citet{Lorenz1963}, for which chaos involves
spiraling orbits surrounding the pair of fixed points (\citet{Sparrow1982}).
Despite the differences described below, it is clear from the calculations
here that chaos only requires this pair with opposite signs of $x_{3}^{*}$.
Moreover, this MCM appears even in the absence of energy conservation
in the gyrostat core. The progression of the three-dimensional orbits
with increasing forcing (Fig. 5) illustrates the dynamics that are
relevant, with this pair of fixed points.

Similar results occur for the model with $\left(F_{1},\epsilon_{3}\right)$
in subclass 2 (SI Fig. 21), subclass 3 (Figure 7), as well as subclass
$4$. Here, there is a pair of fixed points with opposite $x_{3}^{*}$
as before. Analogous plots can be made for $\left(F_{2},\epsilon_{3}\right)$
in subclasses $2$ and $4$. As can be seen from inspecting Table
1, these are the cases where the cubic equations naturally yield a
pair of fixed points with opposite signs of $x_{3}^{*}$. The Jacobian
of the vector fields acquires a skew-symmetric contribution from nonzero
$x_{3}^{*}$ (this presupposes opposite signs of $p$ and $q)$, with
opposite orientations in each hemisphere, giving spirals surrounding
the fixed-point pair that alternate between the hemispheres. This
occurs with energy conservation of the gyrostat core, and it is therefore
not surprising that these cases also are MCMs for the opposite condition
with $p+q+r\neq0$ (Table 1). The effects of increasing the external
forcing for these cases are shown in SI Figs. 34-35 (subclass $2$),
SI Fig. 36 (subclass $3)$, and SI Fig. 39 (subclass $4$). Common
to all these cases is the pair of unstable fixed points with positive
and negative $x_{3}^{*}$.

Thus we can state our first proposition:

\subsubsection*{Proposition 1:}

If the gyrostat core is energy conserving, minimal chaotic models
are governed by the requirement that there are two fixed points with
positive and negative $x_{3}^{*}$.

\subsubsection*{Additional MCMs in the absence of energy conservation of the gyrostat
core}

The additional MCMs found only for the condition of $p+q+r\neq0$
do not fit this pattern. In particular, $\left(F_{3},\epsilon_{2},\epsilon_{3}\right)$
in subclass 2, $\left(F_{2},\epsilon_{3}\right)$ and $\left(F_{3},\epsilon_{3}\right)$
in subclass $3$ (SI Figs. 37-38), and $\left(F_{3},\epsilon_{3}\right)$
in subclass $4$ do not have a pair of fixed points with opposite
$x_{3}^{*}$. In fact, where the forcing occurs on the $x_{3}$ mode,
one cannot expect steady states of both signs and the resulting attractor
looks quite different (SI Fig. 41). This is not limited to coincident
forcing and dissipation, with a similar phenomenon occurring also
with $\left(F_{2},\epsilon_{3}\right)$ of subclass $3$ (SI Fig.
37). The attractors here (e.g., SI Figs. 37-38) are very different
from those of the Lorenz model, which has in its core $p+q=0$ ($r=0$
by the form of streamfunction assumed in the Galerkin projection,
\citet{Hilborn2000}). These attractors with $p+q\neq0$ can resemble
the attractors found in the gyrostat core when there are no constants
of motion (\citet{Seshadri2023}). Sometimes in these cases, for example
$\left(F_{3},\epsilon_{2},\epsilon_{3}\right)$ of subclass 2, and
$\left(F_{3},\epsilon_{3}\right)$ in subclasses $3-4$, forcing and
dissipation appears in the same equation, but this is not always the
case, for example $\left(F_{2},\epsilon_{3}\right)$ in subclass $3$.
Despite their differences, they all depend on the lack of energy conservation
in the gyrostat core, and the absence of the aforementioned pair of
fixed points. It is therefore likely that chaos in these cases is
tied to the loss of invariants in the gyrostat core when $p+q+r\neq0$.
With $p+q+r\neq0$, subclasses 2 and 3 have $1$ quadratic invariant
owing to the presence of two linear feedbacks (\citet{Seshadri2023}),
and the inclusion of forcing might play a role analogous to the third
linear feedback that creates conditions for Hamiltonian chaos in the
gyrostat core. Without energy conservation, subclass $4$ has no quadratic
invariants in the gyrostat core and its MCM $\left(F_{1},\epsilon_{3}\right)$
can encounter very different dynamics on the route to forced and dissipative
chaos as compared to the corresponding MCM of subclass $1$ (SI Fig.
39). While we have not examined the precise role, if any, for the
fixed points in these MCMs, it is clear that these MCMs are tied to
there being fewer quadratic invariants in the gyrostat core. Thus
our second proposition is:

\subsubsection*{Proposition 2:}

If the gyrostat core is not energy conserving, additional minimal
chaotic models can only appear if the gyrostat core has fewer quadratic
invariants as the result of the absence of energy conservation.

\subsubsection*{Broader role of invariant sets}

An important difference between the MCM of subclass $1$ and the Lorenz
model is the absence of the third fixed point at the origin appearing
in the latter, and whose saddle structure permits stable homoclinic
orbits for intermediate values of forcing. The MCM cannot have such
homoclinic orbits, as it possesses only the two fixed points. The
$x_{1}$ axis is invariant in this model, and the effects are seen
in the long stretches of time for which orbits can remain nearby.

To take up the comparison further, consider chaotic cases from subclass
$1$, without the fixed point at the origin but having an invariant
set in the $x_{1}$ axis. When forcing is applied to the first equation,
and dissipation to the second and third equations, as in $\left(F_{1},\epsilon_{3}\right)$
and $\left(F_{1},\epsilon_{2},\epsilon_{3}\right)$, the $x_{1}$
axis remains an invariant set. However there is no fixed point at
the origin. Near this invariant set, the coordinate $x_{1}$ evolves
as $\dot{x_{1}}\approx F_{1}$, growing if $F_{1}>0$. Let us consider
the resulting dynamics near the invariant set by examining the transverse
stability (transverse to this invariant set) of the case $\left(F_{1},\epsilon_{2},\epsilon_{3}\right)$,
given by linearized equations
\begin{equation}
\left\{ \begin{array}{c}
\delta x_{2}'\\
\delta x_{3}'
\end{array}\right\} =\left[\begin{array}{cc}
-\epsilon_{2} & qx_{1}-a\\
a & -\epsilon_{3}
\end{array}\right]\left\{ \begin{array}{c}
\delta x_{2}\\
\delta x_{3}
\end{array}\right\} .
\end{equation}
The above matrix has characteristic equation $\lambda^{2}+\left(\epsilon_{2}+\epsilon_{3}\right)\lambda+\epsilon_{2}\epsilon_{3}-a\left(qx_{1}-a\right)=0$,
with eigenvalues $\lambda=-\epsilon\pm\sqrt{a\left(qx_{1}-a\right)}$,
using $\epsilon_{2}=\epsilon_{3}=\epsilon$. When $a\left(qx_{1}-a\right)<0$,
the eigenvalues are complex conjugate leading to a stable spiral towards
the invariant set. For $0\leq a\left(qx_{1}-a\right)<\epsilon^{2}$,
both eigenvalues are real and negative with local dynamics resembling
a sink. In contrast for $\epsilon^{2}<a\left(qx_{1}-a\right)$ nearby
trajectories are repelled from the invariant set.

Similar results are obtained for the MCM $\left(F_{1},\epsilon_{3}\right)$,
with characteristic polynomial $\lambda^{2}+\epsilon_{3}\lambda-a\left(qx_{1}-a\right)=0$
and eigenvalues
\begin{equation}
\lambda=-\frac{\epsilon_{3}}{2}\pm\sqrt{\frac{\epsilon_{3}^{2}}{4}+a\left(qx_{1}-a\right)}.\label{eq:p28}
\end{equation}
Here points are repelled from the invariant set once $0<a\left(qx_{1}-a\right)$,
and the transition in the neighbourhood of the invariant set remains
the same. These dynamics are shown in Figure 8, for initial condition
$\left(1,0.2,0.2\right)$, fixed $F_{1}=0.021$ so that $x_{1}$ increases
near the invariant set, and $aq>0$ so that the invariant set eventually
becomes unstable. Arrows indicate the direction of the vector field
along the orbit. Near the invariant set the increase of $x_{1}$ is
roughly linear in time, and as it grows the changing stability can
be observed. Initially the invariant set is transverse stable, with
oscillatory dynamics evident from the gaps in the time-series of $x_{2}$
and $x_{3}$ (where these are negative) when plotted on a logarithmic
scale. The logarithmic plots show that $x_{2}$ and $x_{3}$ approach
the invariant set but never reach it, before the set becomes unstable
as $x_{1}$ grows. There is no fixed point along this invariant set
and thus no homoclinic orbits.

Such dynamics has similarities with the Lorenz model
\begin{align}
X' & =-\sigma X+\sigma Y\nonumber \\
Y' & =-XZ+rX-Y\nonumber \\
Z' & =XY-bZ\label{eq:p29}
\end{align}
where $X$ is related to the amplitude of the single mode of the streamfunction
and describes momentum evolution, and $Y,Z$ are related to modes
of temperature evolution. The parameter $\sigma$ is related to dissipation
of momentum through kinematic viscosity (\citet{Hilborn2000}) and,
following our analysis of subclass $1$, it must be nonzero for chaos
in the model. Analogous to the above cases of subclass $1$ there
is an invariant set given by $X=0,Y=0$, with the flow on this set
contracting towards the origin (a fixed point) following $Z'=-bZ$.
The corresponding transverse stability is described by
\begin{equation}
\left\{ \begin{array}{c}
\delta X'\\
\delta Y'
\end{array}\right\} =\left[\begin{array}{cc}
-\sigma & \sigma\\
-Z+r & -1
\end{array}\right]\left\{ \begin{array}{c}
\delta X\\
\delta Y
\end{array}\right\} 
\end{equation}
with the Jacobian matrix having characteristic equation $\lambda^{2}+\left(1+\sigma\right)\lambda+\sigma\left(1-r+Z\right)=0$,
with eigenvalues satisfying $2\lambda=-\left(1+\sigma\right)\pm\sqrt{\left(1+\sigma\right)^{2}-4\sigma\left(1-r+Z\right)}$.
The invariant set repels nearby orbits whenever $\left(1-r+Z\right)<0$.
This gives the well known condition for instability $r>1$ of the
origin $Z=0$, which as a saddle participates in homoclinic orbits.
Thus, despite the differences at the origin, there are important parallels
between the dynamics nearby the respective invariant sets governed
by the changing transverse stability along these sets. In each case,
the common pair of repelling fixed points away from the origin circumscribes
the possibility for chaos.

\pagebreak{}

\begin{landscape}

Table 4: Summary of the major symbols used.

\begin{tabular}{|c|c|}
\hline 
Symbol & Definition\tabularnewline
\hline 
\hline 
$y_{1},y_{2},y_{3}$ & angular velocity of carrier body\tabularnewline
\hline 
$K_{1}^{2},K_{2}^{2},K_{3}^{2}$ & principal moments of inertia of the gyrostat\tabularnewline
\hline 
$h_{1},h_{2},h_{3}$ & angular momentum of the rotor relative to the carrier\tabularnewline
\hline 
$E$ & kinetic energy: $E=\frac{1}{2}\sum_{i=1}^{3}x_{i}^{2}$\tabularnewline
\hline 
$M$ & squared angular momentum: $M=\frac{1}{2}\sum_{i=1}^{3}\left(K_{i}x_{i}+h_{i}\right)^{2}$\tabularnewline
\hline 
$x_{1},x_{2},x_{3}$ & state variables of the transformed gyrostat: $x_{1}=K_{1}y_{1},x_{2}=K_{2}y_{2},x_{3}=K_{3}y_{3}$\tabularnewline
\hline 
$p,q,r$ & quadratic coefficients of transformed gyrostat: $p=K_{2}^{2}-K_{3}^{2},q=K_{3}^{2}-K_{1}^{2},r=K_{1}^{2}-K_{2}^{2}$\tabularnewline
\hline 
$a,b,c$ & linear coefficients of transformed gyrostat: $a=K_{1}h_{1},b=K_{2}h_{2},c=K_{3}h_{3}$\tabularnewline
\hline 
$F_{i}$ & value of constant forcing on the $i$\textsuperscript{th} equation\tabularnewline
\hline 
$\epsilon_{i}$ & coefficient of linear dissipation on the $i$\textsuperscript{th}
equation\tabularnewline
\hline 
$\left(x_{10},x_{20},x_{30}\right)$ & initial condition\tabularnewline
\hline 
$\mathrm{D_{x_{0}}}\left(\mathit{t}\right)\in\mathbb{R}^{\mathit{3\times3}}$ & $\mathrm{D_{x_{0}}^{\mathit{i,j}}\left(\mathit{t}\right)=}\partial x_{i}\left(t\right)/\partial x_{j0}$,
forward sensitivities to perturbation in the initial condition\tabularnewline
\hline 
$\mathrm{M}\left(t\right)$ & $\mathrm{M}\left(t\right)=\mathrm{D_{x_{0}}}\left(t\right)^{T}\mathrm{D_{x_{0}}}\left(t\right)$,
whose spectra yield Lyapunov exponents\tabularnewline
\hline 
$\left(x_{1}^{*},x_{2}^{*},x_{3}^{*}\right)$ & fixed point\tabularnewline
\hline 
$\mathrm{D}_{\mathrm{f}}$ & Jacobian of vector field, usually evaluated at fixed point\tabularnewline
\hline 
$\gamma_{3},\gamma_{2},\gamma_{1},\gamma_{0}$ & coefficients of cubic polynomial giving $x_{3}^{*}$, i.e. $\gamma_{3}x_{3}^{*3}+\gamma_{2}x_{3}^{*2}+\gamma_{1}x_{3}^{*}+\gamma_{0}=0$\tabularnewline
\hline 
$\lambda$ & eigenvalues (of full dynamics, or transverse to invariant subspace,
depending on context)\tabularnewline
\hline 
\end{tabular}

\end{landscape}

\pagebreak{}

\section{Discussion}

This paper is based on large ensemble simulations, to identify the
simplest chaotic models derived from the Volterra gyrostat. This revealed
that minimal chaotic models exist, involving proper subsets of the
forcing and dissipation terms present across the chaotic cases that
are found. For an arbitrary collection of models, the existence of
a minimal chaotic model is not a given. For each of the subclasses
considered here, the existence of MCMs is explicable through common
conditions for chaos in these models.

Our analysis suggests that the forcing and dissipation in these MCMs
together play very specific roles. Both forcing and dissipation must
have specific placement in the equations in order to allow for the
pair of fixed points associated with chaos appearing for the energy
conserving core. Dissipation induces a stable direction in the flow,
which is naturally associated with the linear mode $x_{3}$, owing
to the skew-symmetric property of the nonlinear coupling between $x_{1}$
and $x_{2}$ in each of the subclasses investigated here. Instead,
placing dissipation in $x_{1}$ or $x_{2}$ alone alters the arrangement
of fixed points. Forcing shifts the attractor along the corresponding
axis. To take the example of subclass $1$
\begin{align}
x_{1}' & =px_{2}x_{3}+F_{1}\nonumber \\
x_{2}' & =qx_{3}x_{1}-ax_{3}+F_{2}\nonumber \\
x_{3}' & =ax_{2}-\epsilon_{3}x_{3}+F_{3},\label{eq:p31}
\end{align}
nonzero $\left(F_{3},\epsilon_{3}\right)$ would make $E'=-\epsilon_{3}x_{3}^{2}+F_{3}x_{3}$
and for chaos to be possible $F_{3}x_{3}$ must be positive most of
the time. Such an attractor cannot be distributed across the $x_{1}-x_{2}$
plane, and the coincidence of forcing and dissipation precludes the
fixed point pair associated with the more characteristic MCMs. In
contrast, nonzero $\left(F_{1},\epsilon_{3}\right)$ makes $E'=-\epsilon_{3}x_{3}^{2}+F_{1}x_{1}$
and fixed points on both sides of the $x_{1}-x_{2}$ plane are readily
obtained, giving rise to structures resembling the Lorenz attractor.
The Jacobian of the corresponding fixed-point pair also indicates
how these have opposite orientations in their surrounding flows, which
is crucial for the chaotic orbits that appear.

Identifying MCMs has important applications, even when the underlying
model is not an MCM of the corresponding subclass. For example it
can shed light on the origins of chaos in the model of \citet{Lorenz1963}.
Although nonlinear momentum advection is present in the governing
equations leading to this model, the assumed basis function of streamfunction
renders nonlinear advection's effects absent (\citet{Hilborn2000})
and $X'$ in this model (Eq. (\ref{eq:p29})) has linear terms alone.
Therefore, with two nonlinear terms, and one linearly evolving mode
$X$, which describes momentum evolution, it is not surprising that
chaos in this model requires nonzero kinematic viscosity in order
for momentum dissipation to be present. The model also includes additional
dissipation terms, and it is not the MCM $\left(F_{1},\epsilon_{3}\right)$
but more the case of subclass $1$ with nonzero $\left(F_{1},\epsilon_{2},\epsilon_{3}\right)$
that resembles closely the Lorenz attractor. For the model of \citet{Lorenz1963},
the identification of the MCM for the corresponding subclass (subclass
$1$) clearly establishes that dissipation in $X'$ in Eq. (\ref{eq:p29})
is essential for forced-dissipative chaos, and without the term in
$-\sigma X$ there would be only the single fixed point at the origin.
A further implication is that while Lorenz's model exhibits chaos,
there are further reductions of Lorenz's model that also exhibit chaos.

Furthermore, the Lorenz model has been obtained from a wide variety
of physical processes (\citet{Brindley1980,Gibbon1982,Matson2007}),
and such inquiries can inform the understanding of irregular dynamics
in a variety of systems. Since chaos in these cases does not depend
on the value of $p+q+r$, irregular dynamics can be experienced regardless
of whether energy is conserved. More generally, the identification
of MCMs for the various subclasses can point to a broad range of physical
systems whose modeling assumptions and discretization together give
chaos.

One difference between the Lorenz model and subclass $1$ is that
the former has a third fixed point at the origin, whereas the models
of subclass $1$ have only the pair with nonzero $x_{3}^{*}$. This
is because forcing in the Lorenz model appears through the term $rX$,
with Rayleigh number $r$ being the bifurcation parameter, in contrast
to the case $\left(F_{1},\epsilon_{2},\epsilon_{3}\right)$ of subclass
$1$ having a constant forcing term, leading to only the pair of fixed
points. Yet, the dynamics are quite similar, showing that only this
pair of fixed points is essential to the appearance of chaos. This
also suggests how linear coupling and external forcing can have similar
effects in such models, which is a possible area of further study.

In contrast to the energy conserving core, if the quadratic coefficients
do not sum to zero, there are additional possibilities for forcing
to appear. These possibilities do not require two fixed points with
opposite $x_{3}^{*}$. Moreover, such cases do not admit chaos when
the energy conservation constraint is present in the gyrostat core.
Thus, the appearance of chaos in these cases is closely tied to the
presence of fewer invariants in the gyrostat core if $p+q+r\neq0$.

Chaos in general need not depend on the arrangement of fixed points,
and many simple chaotic models have previously been found that do
not contain any fixed points (\citet{Sprott1994,Jafari2013a}). The
MCMs in this study that only appear when the gyrostat core does not
conserve energy, and for subclasses where this leads to fewer quadratic
invariants, merit further inquiry. Of course, with forcing and dissipation
both present, the model no longer has any quadratic invariants, and
it remains an open question as to whether such chaotic cases arise
from similar pathways as in the gyrostat core without any invariants.
Much of this seems to depend on whether adding forcing has effects
that parallel the linear feedbacks that limit the number of invariants.

There are other subclasses of the Volterra gyrostat, but we have not
considered those with three nonlinear terms, whose fixed-point equations
contain higher degrees. Further generalization of our present results
to such subclasses, as well as to systems of coupled gyrostats, calls
for explicit analysis of these models. Prior studies have indicated
the appearance of Lorenz-like attractors in low-order models of higher
dimension that discretize Rayleigh-Bénard convection with additional
modes (\citet{Musielak2009,Reiterer1998}). Similar attractors are
especially prevalent when such discretization maintains the conservation
properties of these original models. Since such models must contain
systems of coupled gyrostats, similar constraints on fixed points
(and concomitant routes to chaos) might be present in higher dimensions.
Nonlinear feedback is also an important extension of the gyrostat
model (\citet{Lakshmivarahan2008a}). Studying chaos in models involving
coupled gyrostats, investigating the relationships with the number
of invariants, and inquiring into the possibility of a wide range
of chaotic attractors circumscribed by fixed points in these models,
with and without the presence of nonlinear feedbacks, are rich directions
for extension of the analysis contained in this paper.

\section{Conclusions}

The present paper identifies common conditions for minimal chaotic
models (MCMs) across different subclasses of the Volterra gyrostat.
MCMs describe the proper subsets of forcing and dissipation terms
required for chaos to appear and have not been previously investigated.
Our inquiry of MCMs from the Volterra gyrostat showed that when the
gyrostat equations have two nonlinear terms, chaos requires dissipation
of the linear mode. As for forcing, the main factor is whether the
gyrostat core conserves energy. If it does, then there are fewer ways
in which forcing can appear for chaos to be present. Chaos in this
condition requires fixed points with opposite signs of $x_{3}^{*}$
(the linear mode) and this circumscribes where forcing can appear.
The precise results for each subclass are easily found through the
corresponding expression for $x_{3}^{*}$ which takes the form $\gamma_{3}x_{3}^{*2}+\gamma_{1}=0$
for subclasses $1-2$, and with all terms in the cubic equation being
nonzero for subclasses $3-4$. Previous studies have pointed to the
importance of investigating how the arrangement of fixed points can
sometimes circumscribe more complex dynamics (\citet{Eschenazi1989,Gilmore1998}),
and the gyrostat equations present a clear example of this.

Broadly, our findings about forced-dissipative chaos in the Volterra
gyrostat can be summarized as follows. When there is one linear mode
(let us call it $x_{3}$), it sets the direction where points in phase
space experience contraction, and dissipation must necessarily be
present in $x_{3}$. If the placement of external forcing allows two
fixed points with opposite signs of $x_{3}^{*}$, then attractors
that resemble the Lorenz attractor can appear. This condition is necessary
if the gyrostat core is energy-conserving, with these fixed points
acting as repellors. With the energy conserving core, this is the
only way for MCMs to appear. If the gyrostat core does not conserve
energy, then there are further ways for MCMs to arise. These further
arrangements are closely tied to the loss of invariants in subclasses
of the gyrostat having two or more linear feedback terms. Therefore
this possibility is absent from Subclass $1$, whose gyrostat core
maintains two invariants even without energy conservation.

Identifying MCMs has important applications even to systems that are
well understood through simulations. For example, it establishes necessary
conditions for chaos in models that are not MCMs, such as the model
of \citet{Lorenz1963}. Furthermore, the search for MCMs relates the
physics of the governing equations giving rise to chaos to the Galerkin
projected low-order models. When the MCMs arise from the same type
of underlying model, as in the case of the Volterra gyrostat, a common
set of conditions for chaos is possible, with applications to understanding
a wide range of physical systems. Furthermore, in many applications,
distinguishing chaotic and non-chaotic dynamics is critical. Here,
the possibility of identifying MCMs also can bring about a rich set
of design and inverse problems, motivated by having to maintain systems
in their preferred regimes.

\section*{Declarations of interest}

The authors have no competing interests to declare.

\section*{Acknowledgments}

The authors are grateful to S Krishna Kumar, Rajat Masiwal, and two
reviewers for suggesting improvements to the manuscript.

\section*{Appendix 1: Energy conservation of the Volterra gyrostat}

The kinetic energy $E=\frac{1}{2}\sum_{i=1}^{3}x_{i}^{2}$ varies
in time according to
\begin{align*}
E' & =\sum_{i=1}^{3}x_{i}x_{i}'\\
 & =\left(p+q+r\right)x_{1}x_{2}x_{3}+x_{1}\left(bx_{3}-cx_{2}\right)+x_{2}\left(cx_{1}-ax_{3}\right)+x_{3}\left(ax_{2}-bx_{1}\right)\\
 & =\left(p+q+r\right)x_{1}x_{2}x_{3},
\end{align*}
using Eq. (\ref{eq:p3}), where the quadratic terms vanish because
linear feedbacks are skew-symmetric. From the above equation it is
clear that $p+q+r=0$ and skew-symmetry of the linear feedbacks are
both required for energy conservation $E'=0$.

\pagebreak{}

\bibliographystyle{agufull08}
\bibliography{VG_constants}

\pagebreak{}

\begin{landscape}

\begin{figure}
\includegraphics[scale=0.45]{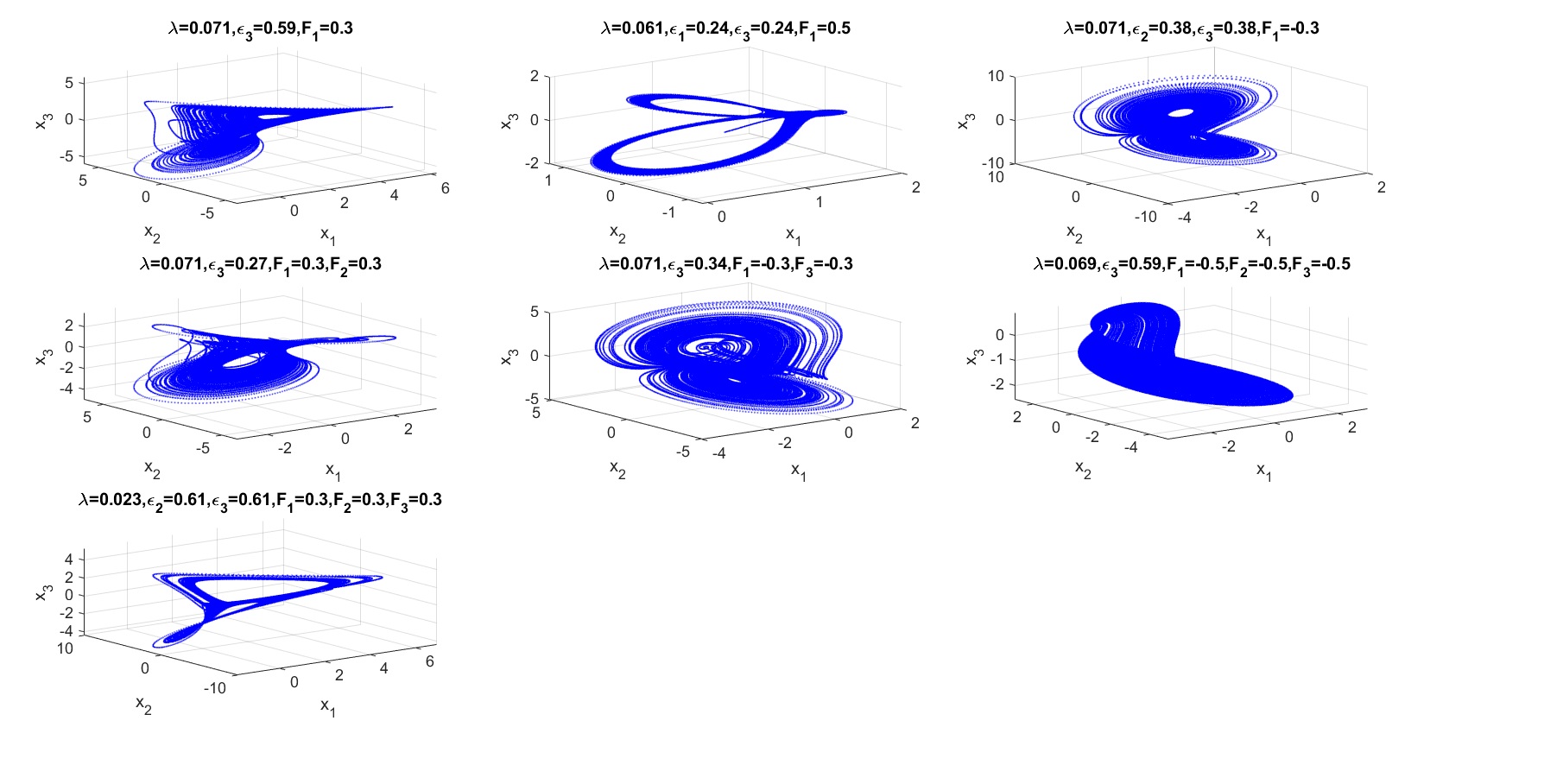}\caption{Orbits of chaotic cases for subclass $1$ with $p+q+r\protect\neq0$:
having nonzero $\left(F_{1},\epsilon_{3}\right)$, $\left(F_{1},\epsilon_{1},\epsilon_{3}\right)$,
$\left(F_{1},\epsilon_{2},\epsilon_{3}\right)$, $\left(F_{1},F_{2},\epsilon_{3}\right)$,
$\left(F_{1},F_{3},\epsilon_{3}\right)$, $\left(F_{1},F_{2},F_{3},\epsilon_{3}\right)$
and $\left(F_{1},F_{2},F_{3},\epsilon_{2},\epsilon_{3}\right)$. Each
of these has nonzero $\left(F_{1},\epsilon_{3}\right)$, the MCM for
subclass $1$.}
\end{figure}

\pagebreak{}

\begin{figure}
\includegraphics[scale=0.5]{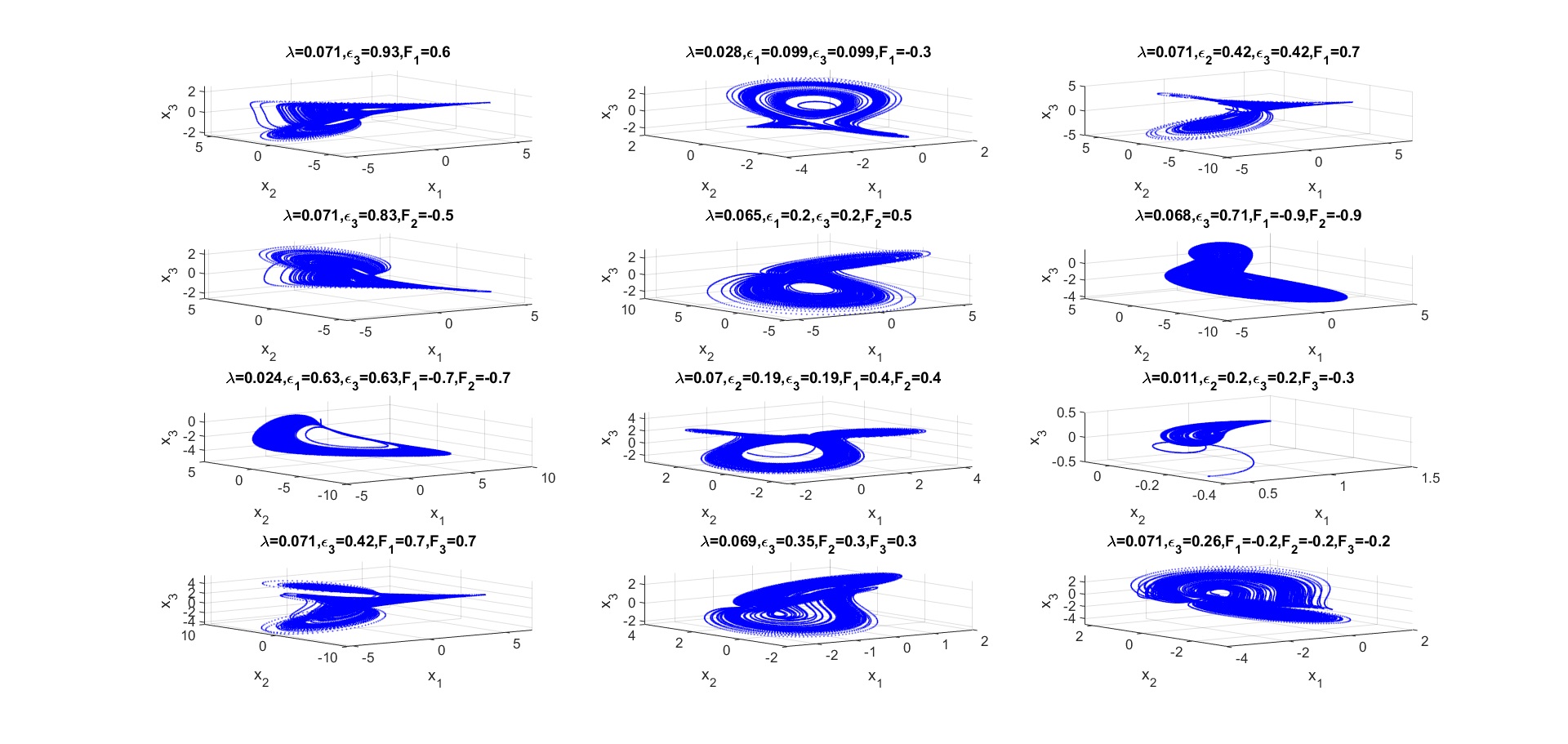}

\caption{Orbits of chaotic cases for subclass $2$ with $p+q+r\protect\neq0$:
having nonzero $\left(F_{1},\epsilon_{3}\right)$, $\left(F_{1},\epsilon_{1},\epsilon_{3}\right)$,
$\left(F_{1},\epsilon_{2},\epsilon_{3}\right)$, $\left(F_{2},\epsilon_{3}\right)$,
$\left(F_{2},\epsilon_{1},\epsilon_{3}\right)$, $\left(F_{1},F_{2},\epsilon_{3}\right)$,
$\left(F_{1},F_{2},\epsilon_{1},\epsilon_{3}\right)$, $\left(F_{1},F_{2},\epsilon_{2},\epsilon_{3}\right)$,
$\left(F_{3},\epsilon_{2},\epsilon_{3}\right)$, $\left(F_{1},F_{3},\epsilon_{3}\right)$,
$\left(F_{2},F_{3},\epsilon_{3}\right)$ and $\left(F_{1},F_{2},F_{3},\epsilon_{3}\right)$.
The simplest chaotic cases, having one forcing and one dissipation
term, involve nonzero $\left(F_{1},\epsilon_{3}\right)$ and $\left(F_{2},\epsilon_{3}\right)$.
Together with $\left(F_{3},\epsilon_{2},\epsilon_{3}\right)$, which
is irreducible, there are $3$ MCMs for subclass $2$.}
\end{figure}

\pagebreak{}

\begin{figure}
\includegraphics[scale=0.5]{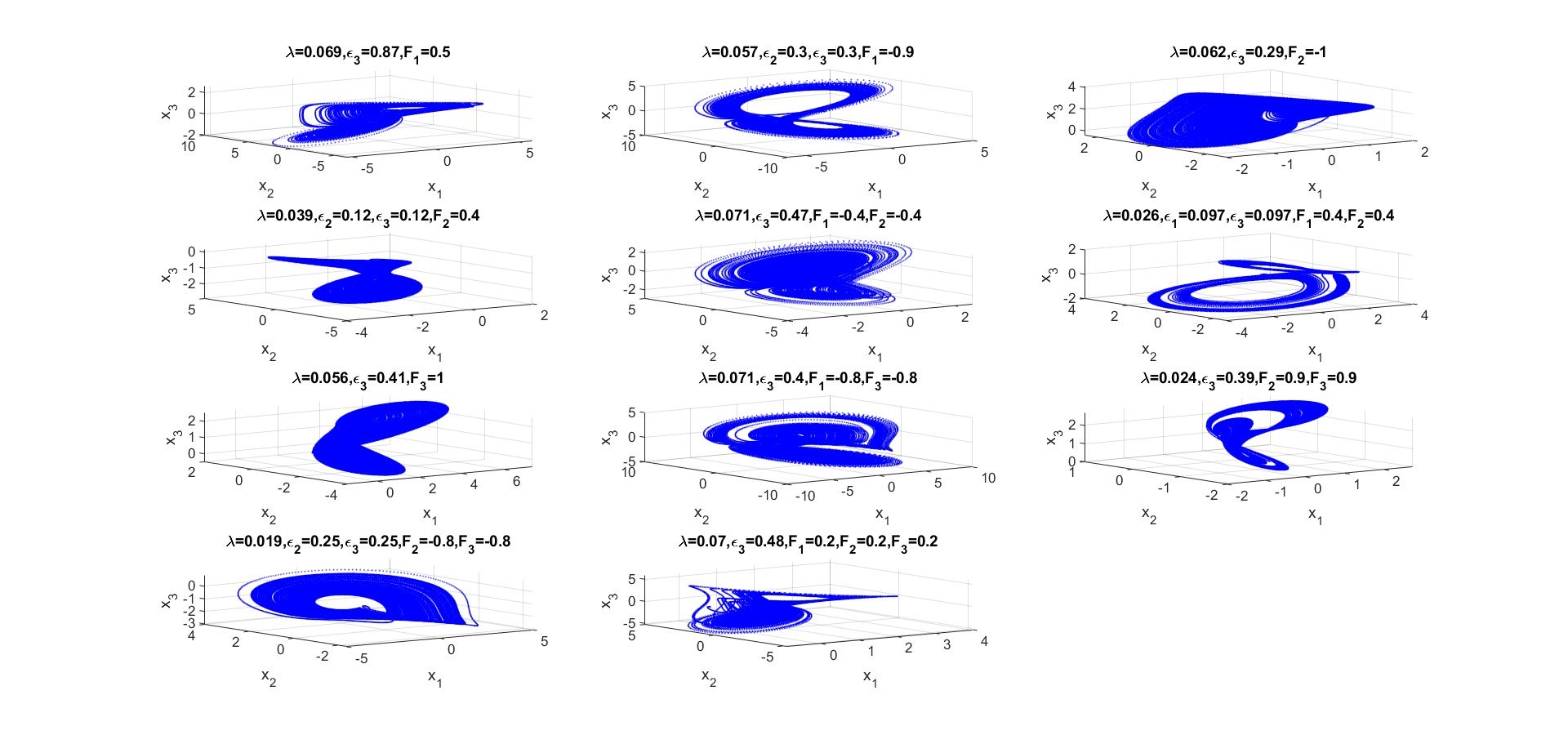}

\caption{Orbits of chaotic cases for subclass $3$ with $p+q+r\protect\neq0$:
having nonzero $\left(F_{1},\epsilon_{3}\right)$, $\left(F_{1},\epsilon_{2},\epsilon_{3}\right)$,
$\left(F_{2},\epsilon_{3}\right)$, $\left(F_{2},\epsilon_{2},\epsilon_{3}\right)$,
$\left(F_{1},F_{2},\epsilon_{3}\right)$, $\left(F_{1},F_{2},\epsilon_{1},\epsilon_{3}\right)$,
$\left(F_{3},\epsilon_{3}\right)$, $\left(F_{1},F_{3},\epsilon_{3}\right)$,
$\left(F_{2},F_{3},\epsilon_{3}\right)$, $\left(F_{2},F_{3},\epsilon_{2},\epsilon_{3}\right)$
and $\left(F_{1},F_{2},F_{3},\epsilon_{3}\right)$. The simplest chaotic
cases, having one forcing and one dissipation term, involve nonzero
$\left(F_{1},\epsilon_{3}\right)$, $\left(F_{2},\epsilon_{3}\right)$,
and $\left(F_{3},\epsilon_{3}\right)$. Each of the chaotic cases
can be reduced to one of these, by setting some terms to zero, and
hence there are $3$ MCMs with nonzero $\left(F_{1},\epsilon_{3}\right)$,
$\left(F_{2},\epsilon_{3}\right)$, and $\left(F_{3},\epsilon_{3}\right)$
for subclass $3$.}
\end{figure}

\pagebreak{}

\begin{figure}
\includegraphics[scale=0.5]{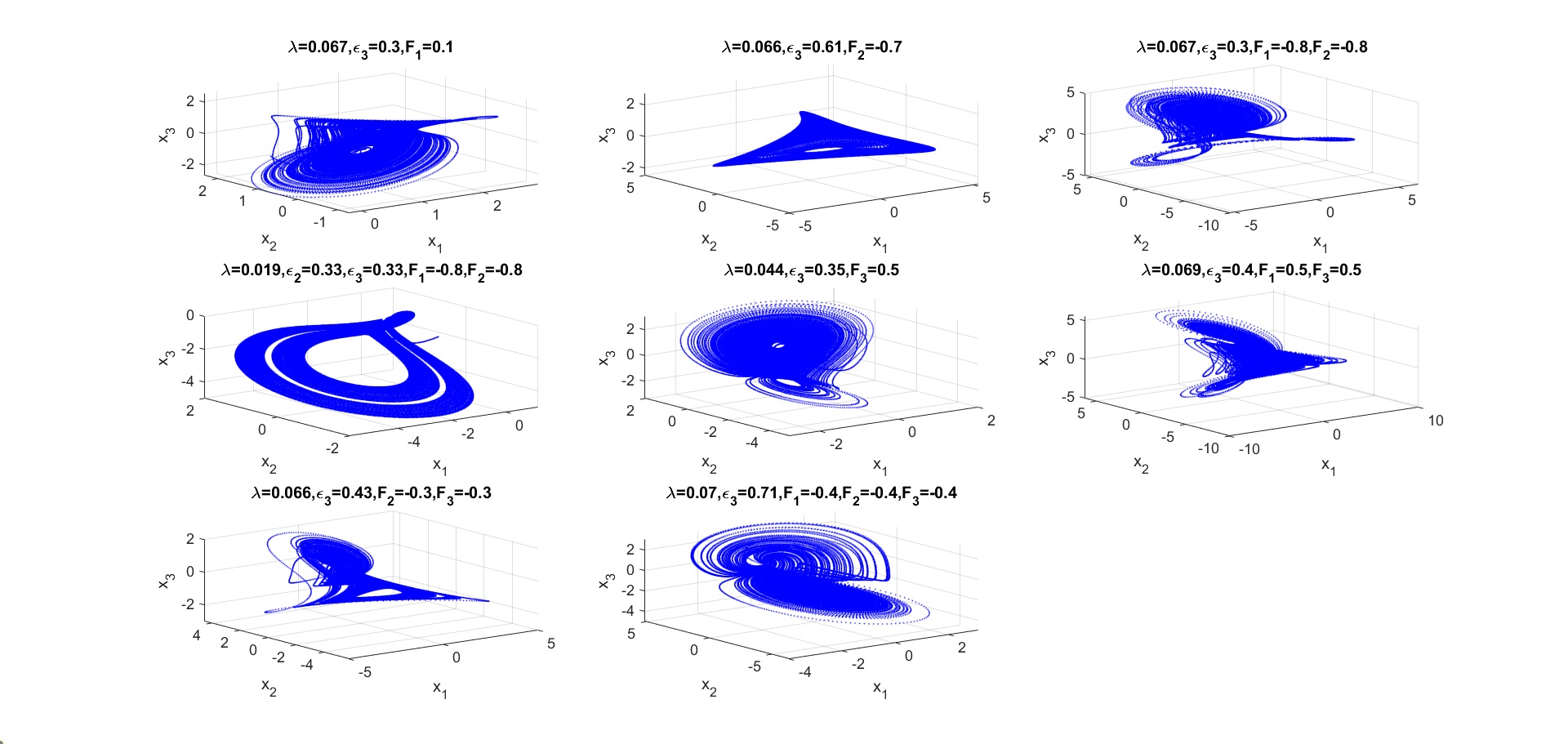}

\caption{Orbits of chaotic cases for subclass $4$ with $p+q+r\protect\neq0$:
having nonzero $\left(F_{1},\epsilon_{3}\right)$, $\left(F_{2},\epsilon_{3}\right)$,
$\left(F_{1},F_{2},\epsilon_{3}\right)$, $\left(F_{1},F_{2},\epsilon_{2},\epsilon_{3}\right)$,
$\left(F_{3},\epsilon_{3}\right)$, $\left(F_{1},F_{3},\epsilon_{3}\right)$,
$\left(F_{2},F_{3},\epsilon_{3}\right)$ and $\left(F_{1},F_{2},F_{3},\epsilon_{3}\right)$.
The simplest chaotic cases, having one forcing and one dissipation
term, involve nonzero $\left(F_{1},\epsilon_{3}\right)$, $\left(F_{2},\epsilon_{3}\right)$,
and $\left(F_{3},\epsilon_{3}\right)$. Each of the chaotic cases
can be reduced to one of these, by setting some terms to zero, and
hence there are $3$ MCMs with nonzero $\left(F_{1},\epsilon_{3}\right)$,
$\left(F_{2},\epsilon_{3}\right)$, and $\left(F_{3},\epsilon_{3}\right)$
for subclass $4$.}
\end{figure}

\pagebreak{}

\begin{figure}
\includegraphics[scale=0.5]{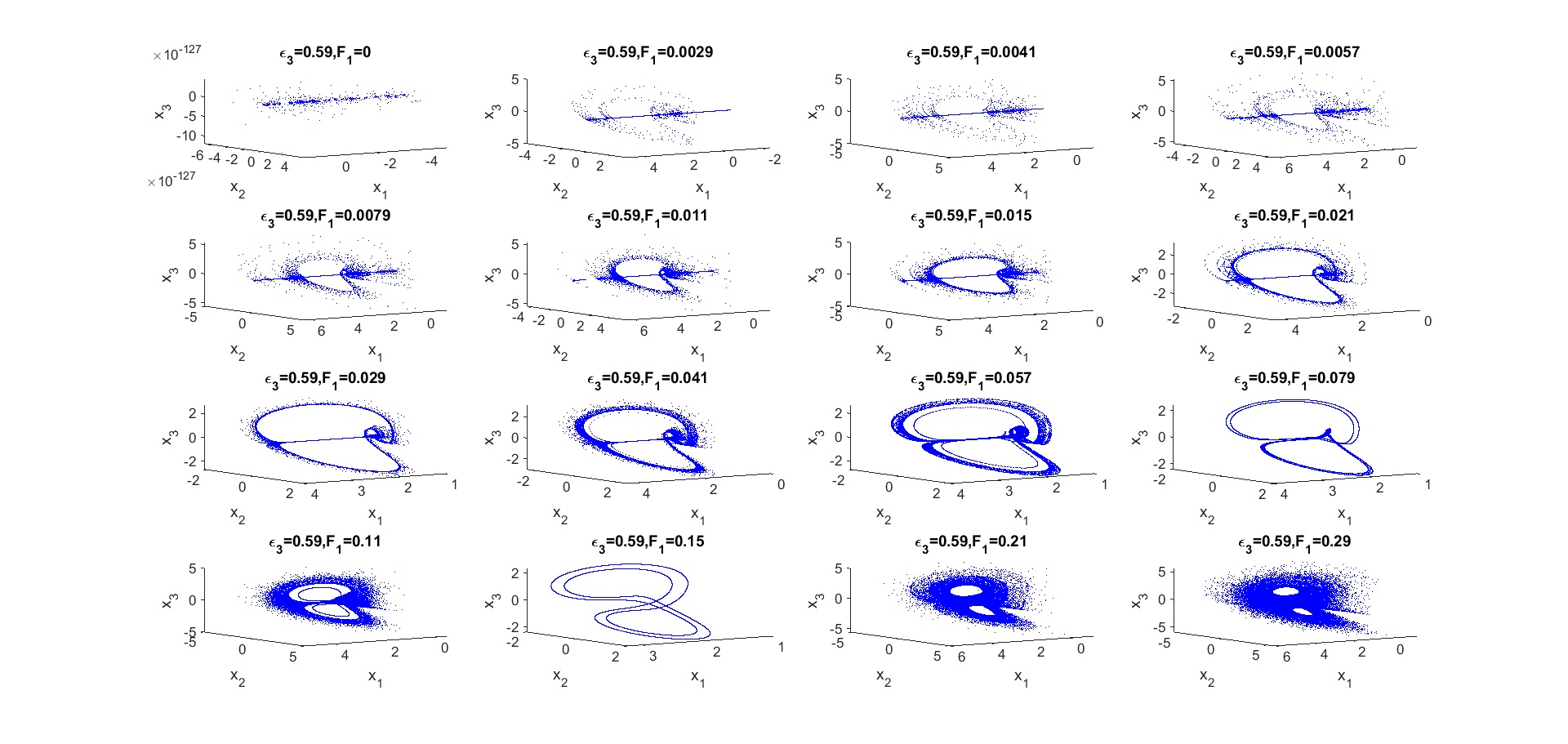}

\caption{Orbits of the MCM $\left(F_{1},\epsilon_{3}\right)$ of subclass $1$,
as the forcing is increased from $0$, for a large initial condition
ensemble. The case with $F_{1}=0.021$ is considered further in Figure
8.}
\end{figure}

\pagebreak{}

\begin{figure}
\includegraphics[scale=0.5]{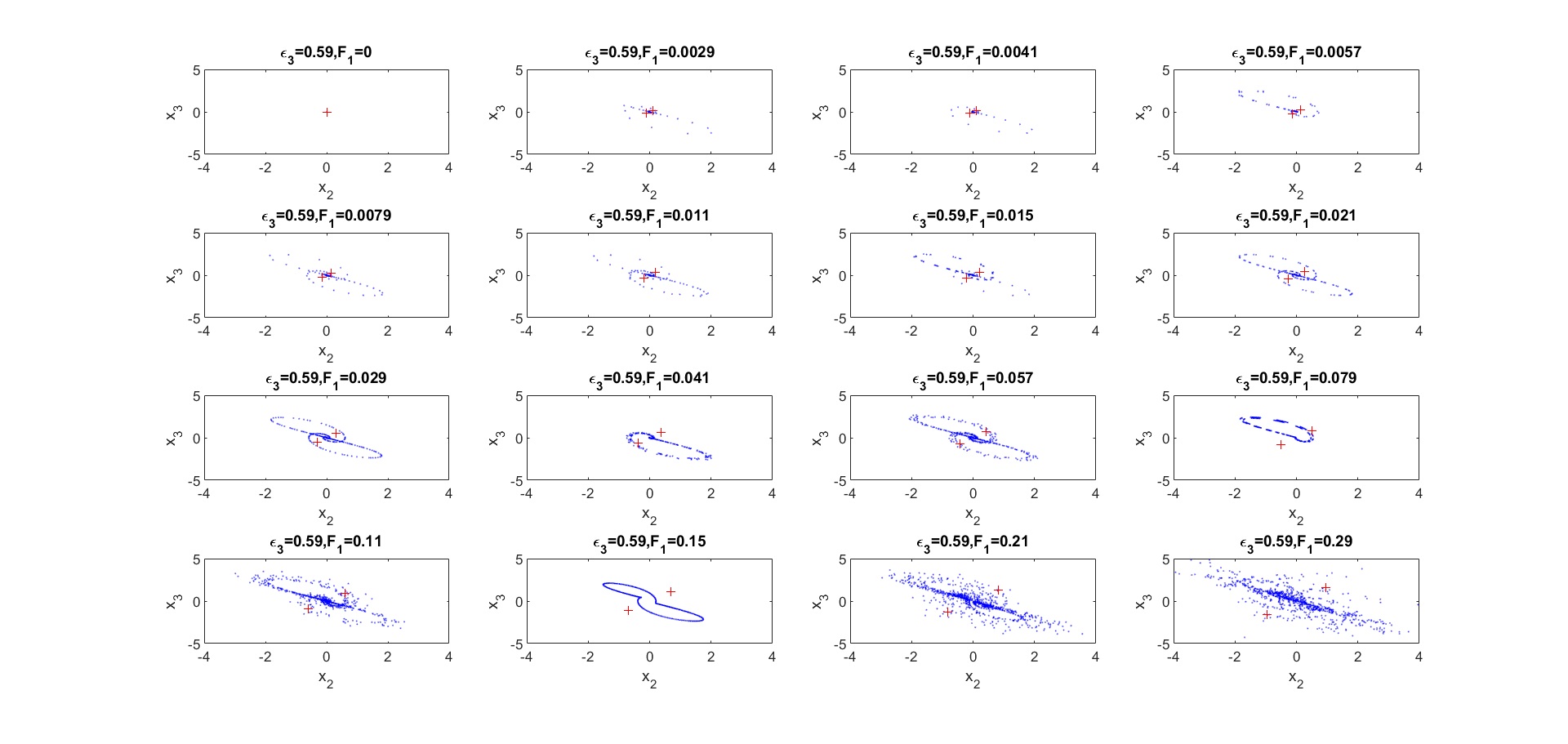}

\caption{Orbits of MCM $\left(F_{1},\epsilon_{3}\right)$ of subclass $1$
in Figure 5 projected onto the $x_{2}-x_{3}$ plane. Red crosses indicate
the fixed points.}
\end{figure}

\pagebreak{}

\begin{figure}
\includegraphics[scale=0.5]{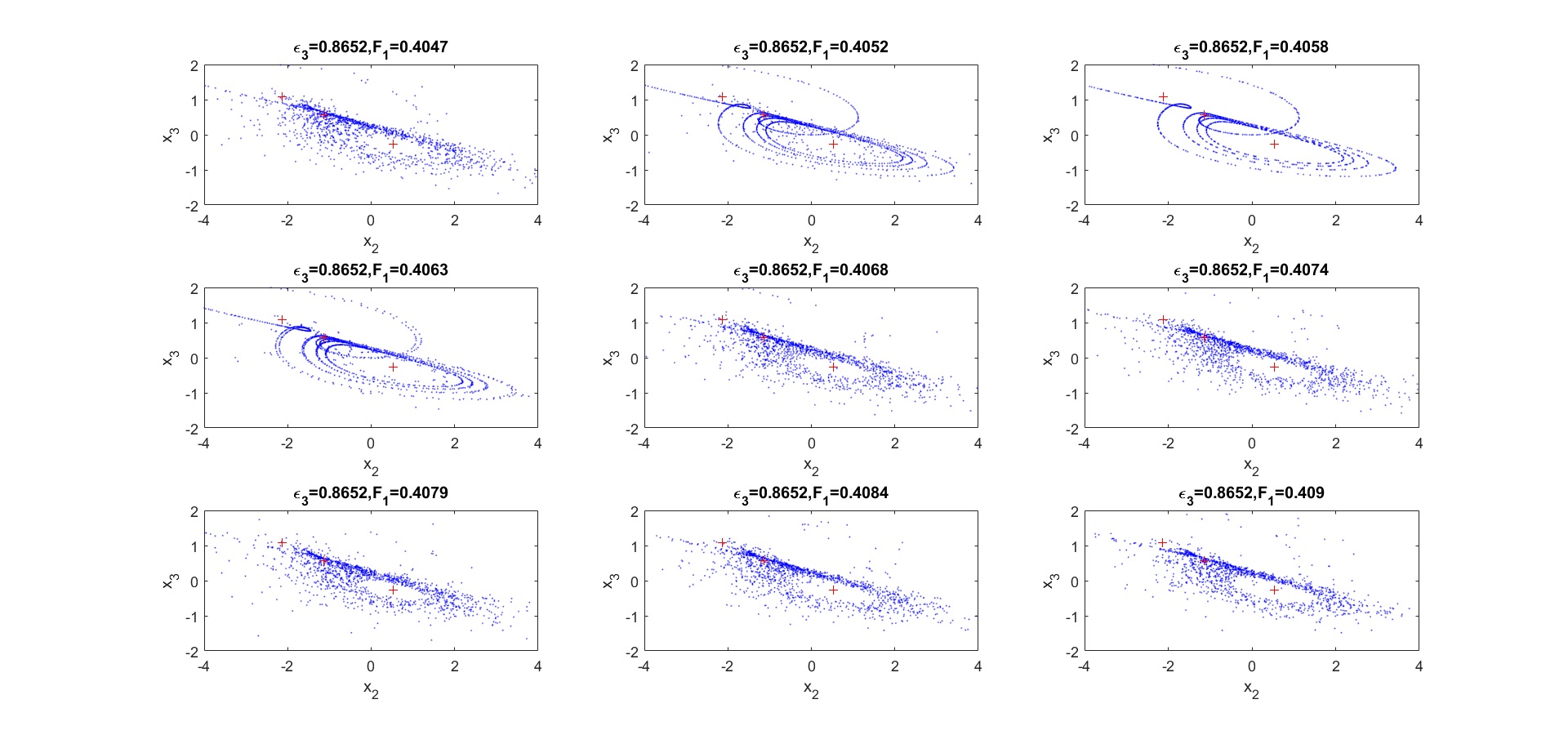}

\caption{Orbits of MCM $\left(F_{1},\epsilon_{3}\right)$ of subclass $3$
projected onto the $x_{2}-x_{3}$ plane, with the value of $F_{1}$
increased over a small interval, for a large initial condition ensemble.
Red crosses indicate the fixed points. The presence of a saddle allows
homoclinic orbits.}
\end{figure}

\pagebreak{}

\begin{figure}
\includegraphics[scale=0.5]{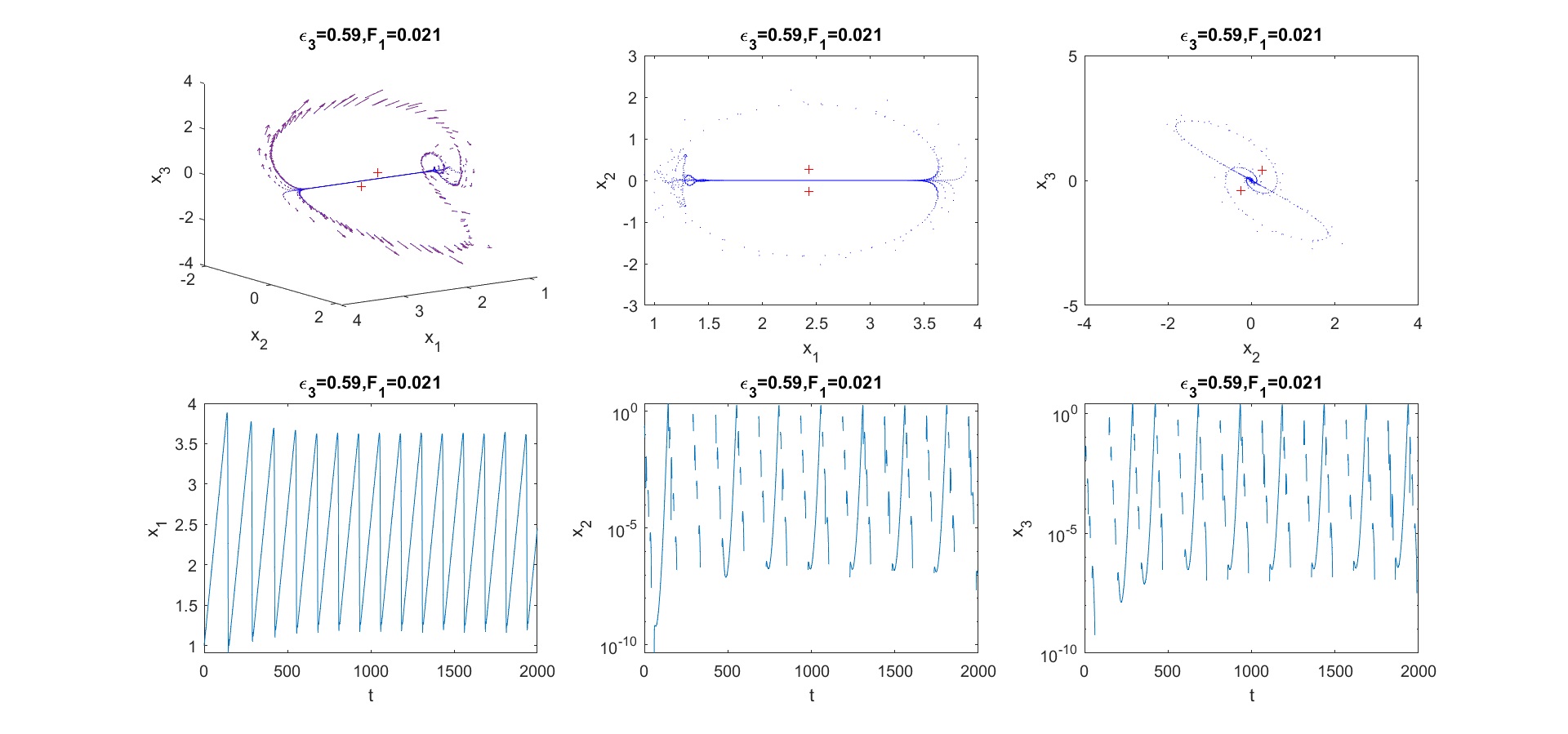}

\caption{Orbit of MCM $\left(F_{1},\epsilon_{3}\right)$ for subclass $1$,
for $F_{1}=0.021$ and the initial condition $\left(1,0.2,0.2\right)$.
Arrows indicate the direction of the vector field along the orbit.
Also shown are corresponding $2-$dimensional projections, and time-series.
The logarithmic scale for the time-series of $x_{2}$ and $x_{3}$
illustrates breaks in the curve where the sign is negative, and the
orbit never reaches the invariant set $x_{2}=0,x_{3}=0$. Thus, the
orbit exhibits oscillatory behavior around points on the invariant
set, which eventually becomes unstable as $x_{1}$ increases. There
is no fixed point along this invariant set, and thus no homoclinic
orbits. The model parameters are $a=-0.94$, $p=0.19$, and $q=-0.39$.}
\end{figure}

\pagebreak{}

\end{landscape}
\end{document}